\documentclass[twocolumn,showpacs,showkeys,preprintnumbers,amssymb,aps,superscriptaddress,prb]{revtex4-2}
\usepackage{graphicx}% Include figure files
\usepackage{dcolumn}% Align table columns on decimal point
\usepackage{color}% Align table columns on decimal point
\usepackage{bm}% bold math

\begin{document}

%\preprint{APS/123-QED}

\title{Unveiling the degeneracy of bound magnon crystals from magnetic and thermodynamic features of the spin-1/2 Heisenberg octahedral chain}
\author{Jozef Stre\v{c}ka}
\email{jozef.strecka@upjs.sk}
\author{Michal Nem\v{c}ík}
\affiliation{Department of Theoretical Physics and Astrophysics, Faculty of Science, P. J. \v{S}af\'{a}rik University, Park Angelinum 9, 040 01 Ko\v{s}ice, Slovakia}

\date{\today}

\begin{abstract}
Magnetic, thermodynamic, and magnetocaloric properties of a spin-$\frac{1}{2}$ Heisenberg octahedral chain with three distinct exchange interactions are investigated in an external magnetic field using the variational method, extended localized-magnon approach, and exact diagonalization. Variational arguments rigorously establish two distinct fragmented phases in the frustrated regime. In the former phase all four spins of each square plaquette form a collective plaquette singlet, whereas in the latter phase two dimer singlets are formed along diagonals of each square plaquette. These bound two-magnon states, supplemented with three localized one-magnon states, enable us to elaborate a generalized localized-magnon theory that is applicable in a frustrated regime across the entire field range as confirmed by comparison with exact diagonalization data. The concept of localized magnons provides a consistent description of low-temperature magnetization curves featuring intermediate one-fifth and three-fifths plateaus, which intersect each other at temperature-independent crossing points determined by the relative degeneracies of competing bound-magnon phases. Field variations of the specific heat reveal a pronounced double-peak structure near each field-driven transition with peak heights depending on the relative degeneracies of the respective bound-magnon states. Our results demonstrate that the system supports highly efficient cooling via adiabatic demagnetization, making it a promising candidate for magnetocaloric refrigeration.
\end{abstract}

\maketitle

\section{Introduction}

Frustrated quantum magnets offer a versatile platform for exploring a broad range of unconventional quantum phenomena emergent at low temperatures including quantum phase transitions, fractional magnetization plateaus regarded as magnetic analogues of the quantum Hall effect, and highly efficient adiabatic demagnetization refrigeration, all of which provide clear signatures of exotic quantum ordered and quantum spin-liquid phases \cite{diep04,lacr11}. Although frustrated magnetism has been a focus of sustained theoretical and experimental interest for nearly half a century since the pioneering work by Toulouse in 1977 introduced the concept of frustration \cite{toul77}, many fundamental aspects still remain unresolved and continue to stimulate intensive research activity in this still highly topical research area. The term spin frustration refers to the inability of a spin system to simultaneously satisfy all pairwise energy minimization constraints either due to competing interactions or specific lattice geometries.

Breakthrough contributions by Johannes Richter have been among the most influential in advancing  the understanding of unconventional quantum features in frustrated Heisenberg spin systems. The detailed analysis of the zero-temperature magnetization curve of a spin-$\frac{1}{2}$ Heisenberg orthogonal-dimer chain by Schulenburg and Richter for instance revealed an infinite series of fractional magnetization plateaus $\frac{n}{2n+2} = \frac{1}{4}, \frac{1}{3}, \cdots, \frac{1}{2}$ ($n$ is positive integer) ranging in between the one-quarter and one-half plateaus \cite{schu02}. To the best of our knowledge, this frustrated quantum spin chain remains the only paradigmatic quantum spin system known to host an infinite sequence of fractional magnetization plateaus. Although no experimental realization of the orthogonal-dimer chain has yet been identified and the remarkable plateau sequence is therefore still awaiting its direct verification, high-field magnetization measurements on SrCu$_2$(BO$_3$)$_2$ affording an experimental realization of a two-dimensional lattice of orthogonal dimers (the so-called Shastry-Sutherland lattice) have confirmed several intriguing fractional magnetization plateaus \cite{kage99,taki13,mats13}. A comprehensive review of magnetization plateaus in several low-dimensional frustrated Heisenberg antiferromagnets was provided by Honecker, Schulenburg, and Richter \cite{hone04}. 

One of the most impactful findings by Richter and coworkers was the discovery of a general route for constructing exact eigenstates of frustrated quantum Heisenberg antiferromagnets from localized one-magnon states \cite{rich02}. These states provide not only a natural explanation for the macroscopic magnetization jump commonly observed in frustrated Heisenberg antiferromagnets at the saturation field, but also reveal the microscopic nature of the many-magnon ground state realized as the last intermediate magnetization plateau prior to saturation as the maximally dense packing of independent bound magnons \cite{rich04,zhit05,derz06,derz07,derz11}. A striking illustration of this concept is the unprecedented series of fractional plateaus detected in ultrahigh-field measurements on Cd-kapellasite CdCu$_3$(OH)$_6$(NO)$_3$ $\cdot$ H$_2$O representing the spin-$\frac{1}{2}$ Heisenberg kagom\'e antiferromagnet, which was successfully interpreted as a consequence of the crystallization of magnons localized on the hexagons of the kagom\'e lattice \cite{okum19}. Moreover, the concept of localized magnons extends far beyond the high-field physics of frustrated Heisenberg antiferromagnets as convincingly evidenced in the comprehensive review on strongly correlated flat-band systems (see Ref. \cite{derz15} and references cited therein). 

Conceptually, flat-band states in frustrated quantum Heisenberg antiferromagnets originate from a destructive quantum interference confining magnons to compact trapping cells of the lattice and hence, one may elaborate a mapping to a classical lattice-gas model capturing many-magnon eigenstates constructed by placing independent bound magnons on separate trapping cells \cite{rich04,zhit05,derz06,derz07,derz11}. A particularly important class of flat-band Heisenberg antiferromagnets comprises those whose Hamiltonians can be rewritten in terms of locally conserved total spins of the trapping cells, because the validity of the localized-magnon theory can be eventually extended to the entire range of magnetic fields rather than only those near saturation. Recent joint collaborations involving among others Johannes Richter and one of the present authors (J.S.) established that a few paradigmatic examples of quantum Heisenberg chains, such octahedral and diamond spin chains, admit a fully consistent description of their magnetic and thermodynamic properties within the extended localized-magnon theory in the highly frustrated regime from zero field up to saturation \cite{stre17,stre18,stre22,karl22}.

In the present article, we adapt the extended localized-magnon theory elaborated in Refs. \cite{stre17,stre18} to a more general version of the spin-$\frac{1}{2}$ Heisenberg octahedral chain, which additionally takes into account the next-nearest-neighbour interaction within the square plaquettes. This extended model enables controlled tuning between two distinct classes of flat-band magnon states. By combining variational arguments with the extended localized-magnon theory, we identify bound-magnon crystal ground states and confirm their stability through the exact diagonalization method.

\section{Heisenberg octahedral chain}
\label{sec:model}

\begin{figure}
\begin{center}
\includegraphics[width=0.49\textwidth]{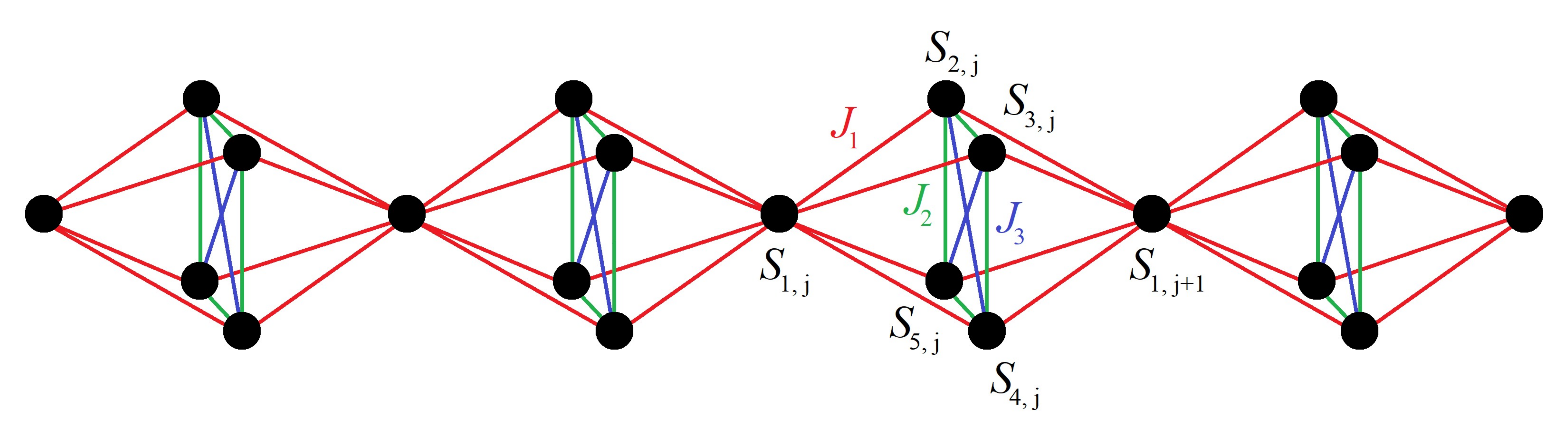}
\end{center}
\vspace{-0.6cm}
\caption{A schematic illustration of the spin-$\frac{1}{2}$ Heisenberg octahedral chain. Red lines represent the monomer-plaquette coupling $J_1$, while green and blue lines correspond to nearest-neighbour and next-nearest-neighbour intra-plaquette couplings $J_2$ and $J_3$, respectively.}
\label{fig1}
\end{figure}

Let us consider a quantum spin-$\frac{1}{2}$ Heisenberg octahedral chain schematically illustrated in Fig.~\ref{fig1} and defined by the Hamiltonian: 
\begin{eqnarray}
\label{ham}
\hat{\cal H} &=& 
\sum_{j=1}^{N} \Bigl[ J_1 (\boldsymbol{\hat{S}}_{1,j} + \boldsymbol{\hat{S}}_{1,j+1}) \!\cdot\! (\boldsymbol{\hat{S}}_{2,j} + \boldsymbol{\hat{S}}_{3,j} + \boldsymbol{\hat{S}}_{4,j} + \boldsymbol{\hat{S}}_{5,j}) \Bigr.  \nonumber \\
&+& J_2 (\boldsymbol{\hat{S}}_{2,j}\!\cdot\!\boldsymbol{\hat{S}}_{3,j} + \boldsymbol{\hat{S}}_{3,j}\!\cdot\!\boldsymbol{\hat{S}}_{4,j}
+ \boldsymbol{\hat{S}}_{4,j}\!\cdot\!\boldsymbol{\hat{S}}_{5,j} + \boldsymbol{\hat{S}}_{5,j}\!\cdot\!\boldsymbol{\hat{S}}_{2,j}) \nonumber \\
\Bigl. &+& J_3 (\boldsymbol{\hat{S}}_{2,j}\!\cdot\!\boldsymbol{\hat{S}}_{4,j} + \boldsymbol{\hat{S}}_{3,j}\!\cdot\!\boldsymbol{\hat{S}}_{5,j}) -h \sum_{i=1}^{5} \hat{S}_{i,j}^{z} \Bigr].
\end{eqnarray}
Here, $\boldsymbol{\hat{S}}_{i,j} \equiv (\hat{S}_{i,j}^x, \hat{S}_{i,j}^y, \hat{S}_{i,j}^z)$ denotes a spin-$\frac{1}{2}$ operator assigned to a lattice site uniquely identified by two subscripts: the former index $i$ specifies the position within a unit cell and the latter index $j$ labels the unit cell itself. Each unit cell involves one monomeric spin ($i=1$) and four spins belonging to a square plaquette ($i=2-5$). The coupling constant $J_1$ represents the monomer-plaquette interaction between nearest-neighbor spins from monomeric and plaquette sites, while the coupling constants $J_2$ and $J_3$ denote the nearest-neighbour and next-nearest-neighbour interactions within square plaquettes. The Zeeman term with field strength $h \geq 0$ accounts for the magnetostatic energy of the spins in an external magnetic field. For convenience, periodic boundary conditions $\boldsymbol{S}_{1,N+1} \equiv \boldsymbol{S}_{1,1}$ are imposed. 

\subsection{Fragmented bound-magnon crystals from the variational method}
\label{vm} 

Exact ground states of the spin-$\frac{1}{2}$ Heisenberg octahedral chain in the highly frustrated regime and under sufficiently low magnetic fields can be rigorously determined using the variational principle \cite{shas81,bose89,bose90,bose92}. To this end, the total Hamiltonian (\ref{ham}) can be decomposed into a sum of cell Hamiltonians: 
\begin{eqnarray}
\hat{\cal H} = \sum_{j=1}^{N} \sum_{k=0}^{1} \hat{\cal H}_{j,k},
\label{hamsum}
\end{eqnarray}
where each cell Hamiltonian $\hat{\cal H}_{j,k}$ includes  all interaction terms of an elementary five-spin cluster forming a square pyramid: 
\begin{eqnarray}
\hat{\cal H}_{j,k} &=& J_1 \boldsymbol{\hat{S}}_{1,j+k} \cdot (\boldsymbol{\hat{S}}_{2,j} + \boldsymbol{\hat{S}}_{3,j} 
                                 + \boldsymbol{\hat{S}}_{4,j} + \boldsymbol{\hat{S}}_{5,j}) \nonumber \\
                  &+&  \frac{J_2}{2} (\boldsymbol{\hat{S}}_{2,j}\!\cdot\!\boldsymbol{\hat{S}}_{3,j} + \boldsymbol{\hat{S}}_{3,j}\!\cdot\!\boldsymbol{\hat{S}}_{4,j}
+ \boldsymbol{\hat{S}}_{4,j}\!\cdot\!\boldsymbol{\hat{S}}_{5,j} + \boldsymbol{\hat{S}}_{5,j}\!\cdot\!\boldsymbol{\hat{S}}_{2,j}) \nonumber \\
 &+& \frac{J_3}{2} (\boldsymbol{\hat{S}}_{2,j}\!\cdot\!\boldsymbol{\hat{S}}_{4,j} + \boldsymbol{\hat{S}}_{3,j}\!\cdot\!\boldsymbol{\hat{S}}_{5,j}) - \frac{h}{2} \Bigl(\! \hat{S}_{1,j+k}^z \!+\! \sum_{i=2}^{5} \! \hat{S}_{i,j}^{z}\Bigr).
\nonumber
\end{eqnarray}

The factors of $\frac{1}{2}$ in front of the Zeeman term $h$ and the intra-plaquette interactions $J_2$ and $J_3$ eliminate their double counting as these contributions are symmetrically shared between two consecutive cell Hamiltonians. The variational principle provides a rigorous lower bound for the ground-state energy $E_0$ of the spin-$\frac{1}{2}$ Heisenberg octahedral chain:
\begin{eqnarray}
E_0 \!=\! \langle \Psi_0 | \hat{\cal H} | \Psi_0 \rangle \!=\! \langle \Psi_0 | \! \sum_{j=1}^{N} \! \sum_{k=0}^{1} \! \hat{\cal H}_{j,k} | \Psi_0 \rangle \!\geq\! \sum_{j=1}^{N} \! \sum_{k=0}^{1} \! \varepsilon_{j,k}^0, 
\label{var}
\end{eqnarray}
since the global ground-state eigenvector $| \Psi_0 \rangle$ alternatively serves as a variational trial function for each local five-spin cluster. The inequality (\ref{var}) implies that the total ground-state energy $E_0$ cannot be smaller than the sum of the lowest eigenenergies  $\varepsilon_{j,k}^0$ of the individual clusters. The energy spectrum of a five-spin cluster can be expressed in terms of five quantum spin numbers:
\begin{eqnarray}
\varepsilon_{j,k} &=& \frac{J_1}{2} [S_{T,j,k}(S_{T,j,k}+1) - S_{\square,j} (S_{\square,j} + 1)] \nonumber \\
    &-& \frac{J_2}{4} [S_{24,j} (S_{24,j} + 1) + S_{35,j} (S_{35,j} + 1)] \nonumber \\
    &+& \frac{J_3}{4} [S_{24,j} (S_{24,j} + 1) + S_{35,j} (S_{35,j} + 1)] \nonumber \\
    &-& \frac{3}{8} J_1 + \frac{J_2}{4} S_{\square,j} (S_{\square,j} + 1) - \frac{3}{4} J_3- h S_{T,j,k}^{z},
\label{spec}
\end{eqnarray}
which determine the total spin of the square-pyramidal cluster $S_{T,j,k}$ and its $z$-component $S_{T,j,k}^{z}$, the total spin of the square plaquette $S_{\square,j}$ and the total spin of two spin pairs from opposite corners of a square plaquette $S_{24,j}$ and $S_{35,j}$, respectively.

It follows from Eq. (\ref{spec}) that the lowest-energy eigenstate of the five-spin cluster in the parameter region $J_2>2J_1$, $J_3<J_2$, and $h<J_1+J_2$ is a doublet characterized by the quantum spin numbers $S_{T,j,k} = |S_{T,j,k}^{z}| = \frac{1}{2}$, $S_{\square,j} = 0$, $S_{24,j}= 1$ and $S_{35,j} = 1$. In this state, the monomer spins are completely independent (decoupled) from the four spins forming the square plaquette constituting a collective plaquette-singlet state:
\begin{eqnarray}
|0,1,1\rangle_j &=& |S_{\square,j}=0, S_{24,j} = 1, S_{35,j}=1 \rangle   \nonumber \\
&=& \!\frac{1}{\sqrt{3}}(|\!\!\uparrow_{2,j}\downarrow_{3,j}\uparrow_{4,j}\downarrow_{5,j}\rangle + |\!\!\downarrow_{2,j}\uparrow_{3,j}\downarrow_{4,j}\uparrow_{5,j}\rangle)  \nonumber \\
&-& \frac{1}{\sqrt{12}} (|\!\!\uparrow_{2,j}\uparrow_{3,j}\downarrow_{4,j}\downarrow_{5,j}\rangle + |\!\!\uparrow_{2,j}\downarrow_{3,j}\downarrow_{4,j}\uparrow_{5,j}\rangle \nonumber \\
&+& |\!\!\downarrow_{2,j}\uparrow_{3,j}\uparrow_{4,j}\downarrow_{5,j}\rangle + |\!\!\downarrow_{2,j}\downarrow_{3,j}\uparrow_{4,j}\uparrow_{5,j}\rangle),\!
\label{PS}
\end{eqnarray}
which may be alternatively interpreted as a bound two-magnon state. Extending this local eigenstate to the entire spin-$\frac{1}{2}$ Heisenberg octahedral chain results in a fragmented ground state referred to as the bound two-magnon plaquette (TMP) phase:  
\begin{eqnarray}
|{\rm TMP} \rangle \!\!=\!\! \prod_{j=1}^N \! |\!\!\uparrow_{1,j}\rangle \!&\otimes&\! 
\Bigl[\!\frac{1}{\sqrt{3}}(|\!\!\uparrow_{2,j}\downarrow_{3,j}\uparrow_{4,j}\downarrow_{5,j}\rangle \!\!+\!\! |\!\!\downarrow_{2,j}\uparrow_{3,j}\downarrow_{4,j}\uparrow_{5,j}\rangle)  \nonumber \\
\!&-&\! \frac{1}{\sqrt{12}} (|\!\!\uparrow_{2,j}\uparrow_{3,j}\downarrow_{4,j}\downarrow_{5,j}\rangle \!\!+\!\! |\!\!\uparrow_{2,j}\downarrow_{3,j}\downarrow_{4,j}\uparrow_{5,j}\rangle \nonumber \\
\!&+&\! |\!\!\downarrow_{2,j}\uparrow_{3,j}\uparrow_{4,j}\downarrow_{5,j}\rangle \!\!+\!\! |\!\!\downarrow_{2,j}\downarrow_{3,j}\uparrow_{4,j}\uparrow_{5,j}\rangle) \Bigr]\!. \nonumber \\
\label{TMP}
\end{eqnarray}
The TMP phase thus constitutes an exact ground state of the spin-$\frac{1}{2}$ Heisenberg octahedral chain  in the parameter range $J_2>2J_1$, $J_3<J_2$, and $h<J_1+J_2$. 

Within the parameter regime $J_3>2J_1$, $J_3>J_2$, and $h<J_1+J_3$, the lowest-energy eigenstate of the five-spin cluster in the energy spectrum (\ref{spec}) is another doublet state with quantum spin numbers $S_{T,j,k} = |S_{T,j,k}^{z}| = \frac{1}{2}$, $S_{\square,j} = 0$, $S_{24,j}= 0$ and $S_{35,j} = 0$. In this particular case, the four plaquette spins constitute a direct product of two dimer singlets formed along diagonals of a square plaquette:
\begin{eqnarray}
|0\!\!&,&\!\!0,0\rangle_j \!=\! |S_{\square,j}=0, S_{24,j} = 0, S_{35,j}=0 \rangle   \nonumber \\
&=& \!\frac{1}{\sqrt{2}}(|\!\!\uparrow_{2,j}\downarrow_{4,j}\rangle 
                       \!-\! |\!\!\downarrow_{2,j}\uparrow_{4,j}\rangle)  
											\!\otimes\! \frac{1}{\sqrt{2}} (|\!\!\uparrow_{3,j}\downarrow_{5,j}\rangle 
                            \!-\! |\!\!\downarrow_{3,j}\uparrow_{5,j}\rangle \nonumber \\
						&=& \!\frac{1}{2}(|\!\!\uparrow_{2,j}\uparrow_{3,j}\downarrow_{4,j}\downarrow_{5,j}\rangle	
													   -|\!\!\uparrow_{2,j}\downarrow_{3,j}\downarrow_{4,j}\uparrow_{5,j}\rangle \nonumber \\
						&&					 -|\!\!\downarrow_{2,j}\uparrow_{3,j}\uparrow_{4,j}\downarrow_{5,j}\rangle 
														 +|\!\!\downarrow_{2,j}\downarrow_{3,j}\uparrow_{4,j}\uparrow_{5,j}\rangle). 
\label{DS}
\end{eqnarray}
The formation of two dimer-singlet states provides another bound two-magnon state, in which the plaquette spins are completely decoupled from the monomer spins. Extending this local eigenstate to the entire spin-$\frac{1}{2}$ Heisenberg octahedral chain affords another fragmented ground state referred to as the bound two-magnon dimer (TMD) phase:  
\begin{eqnarray}
|{\rm TMD} \rangle \!=\! \prod_{j=1}^N \! |\!\!\uparrow_{1,j}\rangle \!&\otimes&\! 
 \!\frac{1}{\sqrt{2}}(|\!\!\uparrow_{2,j}\downarrow_{4,j}\rangle - |\!\!\downarrow_{2,j}\uparrow_{4,j}\rangle)  \nonumber \\
\!&\otimes&\! \frac{1}{\sqrt{2}} (|\!\!\uparrow_{3,j}\downarrow_{5,j}\rangle - |\!\!\downarrow_{3,j}\uparrow_{5,j}\rangle). 
\label{TMD}
\end{eqnarray}
It can be proved that the TMD phase represents an exact ground state of the spin-$\frac{1}{2}$ Heisenberg octahedral chain within the parameter region $J_3>2J_1$, $J_3>J_2$, and $h<J_1+J_3$. It could be concluded that the spin-$\frac{1}{2}$ Heisenberg octahedral chain supports in a highly frustrated regime two distinct fragmented TMP and TMD ground states, where monomeric spins are fully aligned with the external magnetic field due to the formation of either a collective plaquette-singlet or a product of dimer-singlet states on square plaquettes. Owing to the complete polarization of the monomer spins, both fragmented TMP and TMD ground states give rise to an intermediate one-fifth plateau in the zero-temperature magnetization curve.

\subsection{Bound many-magnon crystals from exact one-magnon states}
\label{lmgs}

At sufficiently high magnetic fields exceeding the saturation value, the lowest-energy eigenstate of the spin-$\frac{1}{2}$ Heisenberg octahedral chain is the fully polarized ferromagnetic (FM) state $|{\rm FM} \rangle = \prod_{j=1}^N |\!\!\uparrow_{1,j}\uparrow_{2,j}\uparrow_{3,j}\uparrow_{4,j}\uparrow_{5,j}\rangle$ with the corresponding energy eigenvalue $E_{\rm FM} = E_{\rm FM}^0 - \frac{5N}{2} h$, where $E_{\rm FM}^0 = N (2J_1 + J_2 + \frac{J_3}{2})$ denotes the respective zero-field contribution. In what follows, we will exactly calculate one-magnon eigenstates allowing rigorous determination of exact ground states emerging just below the saturation field. To this end, the one-magnon eigenstates are constructed within the orthonormal basis set $|l, j \rangle = \hat{S}_{l,j}^{-} |{\rm FM} \rangle$ ($l=1-5$, $j=1-N$), which spans the sector $S_T^z = \frac{5N}{2} - 1$ with a single spin deviation relative to the fully polarized FM state. Applying the zero-field part of the Hamiltonian (\ref{ham}) to this basis yields the following system of coupled equations:
\begin{eqnarray}
\hat{\cal H} |1, j\rangle &=& (E_{\rm FM}^0 - 4 J_1) |1, j\rangle  + \frac{J_1}{2} \sum_{i=2}^5 (|i, j-1\rangle + |i, j\rangle), \nonumber \\
\hat{\cal H} |2, j\rangle  &=& (E_{\rm FM}^0 - J_1 - J_2 - \frac{J_3}{2}) |2, j\rangle + \frac{J_3}{2} |4, j\rangle \nonumber \\
&+& \frac{J_2}{2} (|3, j\rangle + |5, j\rangle) + \frac{J_1}{2} (|1, j\rangle + |1, j+1\rangle), \nonumber \\
\hat{\cal H} |3, j\rangle  &=& (E_{\rm FM}^0 - J_1 - J_2 - \frac{J_3}{2}) |3, j\rangle  + \frac{J_3}{2} |5, j\rangle \nonumber \\
&+& \frac{J_2}{2} (|2, j\rangle + |4, j\rangle) + \frac{J_1}{2} (|1, j\rangle + |1, j+1\rangle), \nonumber \\
\hat{\cal H} |4, j\rangle  &=& (E_{\rm FM}^0 - J_1 - J_2 - \frac{J_3}{2}) |4, j\rangle  + \frac{J_3}{2} |2, j\rangle \nonumber \\
&+& \frac{J_2}{2} (|3, j\rangle + |5, j\rangle) + \frac{J_1}{2} (|1, j\rangle + |1, j+1\rangle), \nonumber \\
\hat{\cal H} |5, j\rangle  &=& (E_{\rm FM}^0 - J_1 - J_2 - \frac{J_3}{2}) |5, j\rangle  + \frac{J_3}{2} |3, j\rangle \nonumber \\
&+& \frac{J_2}{2} (|2, j\rangle + |4, j\rangle) + \frac{J_1}{2} (|1, j\rangle + |1, j+1\rangle), 
\label{lm}
\end{eqnarray}
which serves to solve the one-magnon eigenvalue problem $\hat{\cal H} |\Psi_k\rangle = E_k^0 |\Psi_k\rangle$ in a zero field by adopting the Bloch spin waves $|\Psi_k\rangle = \sum_{l=1}^{5} \sum_{j=1}^{N} c_{l} {\rm e}^{{\rm i} k j}|l, j \rangle$ with the momentum $k$ defined by the periodic boundary conditions. Substituting this form into the linear system of five coupled equations (\ref{lm}) for the probability amplitudes $c_{i,k}$ leads to the characteristic equation: 
\begin{widetext}
\begin{eqnarray}
\left|
\begin{array}{ccccc}
-4J_1-\varepsilon_k & \frac{J_1}{2} (1 + {\rm e}^{-{\rm i} k}) & \frac{J_1}{2} (1 + {\rm e}^{-{\rm i} \kappa}) 
& \frac{J_1}{2} (1 + {\rm e}^{-{\rm i} k}) & \frac{J_1}{2} (1 + {\rm e}^{-{\rm i} \kappa}) \\[4pt] 	
\frac{J_1}{2} (1 + {\rm e}^{{\rm i} k}) & -J_1-J_2-\frac{J_3}{2}-\varepsilon_k & \frac{J_2}{2} & \frac{J_3}{2} & \frac{J_2}{2} \\[4pt]
\frac{J_1}{2} (1 + {\rm e}^{{\rm i} k}) & \frac{J_2}{2} & -J_1-J_2-\frac{J_3}{2}-\varepsilon_k & \frac{J_2}{2} & \frac{J_3}{2} \\[4pt]
\frac{J_1}{2} (1 + {\rm e}^{{\rm i} k}) & \frac{J_3}{2} & \frac{J_2}{2} & -J_1-J_2-\frac{J_3}{2}-\varepsilon_k & \frac{J_2}{2} \\[4pt]
\frac{J_1}{2} (1 + {\rm e}^{{\rm i} k}) & \frac{J_2}{2} & \frac{J_3}{2} & \frac{J_2}{2} & -J_1-J_2-\frac{J_3}{2}-\varepsilon_k \\
\end{array}
\right| = 0,
\label{det}
\end{eqnarray}
\end{widetext}
where $\varepsilon_k = E_k^0 - E_{\rm FM}^0$ denotes the energy difference between the one-magnon state and the fully polarized FM state in a zero magnetic field. The one-magnon energy spectrum of the spin-$\frac{1}{2}$ Heisenberg octahedral chain in a zero field is composed of five distinct bands:
\begin{eqnarray}
\varepsilon_{1}   &=& - J_1 - 2J_2,  \nonumber \\
\varepsilon_{2,3} &=& - J_1 - J_2 - J_3,  \nonumber \\
\varepsilon_{4,5} &=& - \frac{J_1}{2} \left(5 \pm \sqrt{17 + 8 \cos k} \right).  
\label{oms}
\end{eqnarray}
These five one-magnon energy bands are illustrated in Fig.~\ref{fig3} for a few representative values of the interaction ratio $J_2/J_1$ and $J_2/J_1$. It is worth emphasizing that three out of five one-magnon bands (\ref{oms}) are completely dispersionless (flat), which indicates the existence of localized magnons confined within those flat bands \cite{derz06,derz15}. 

\begin{figure}
\begin{center}
\includegraphics[width=0.48\textwidth]{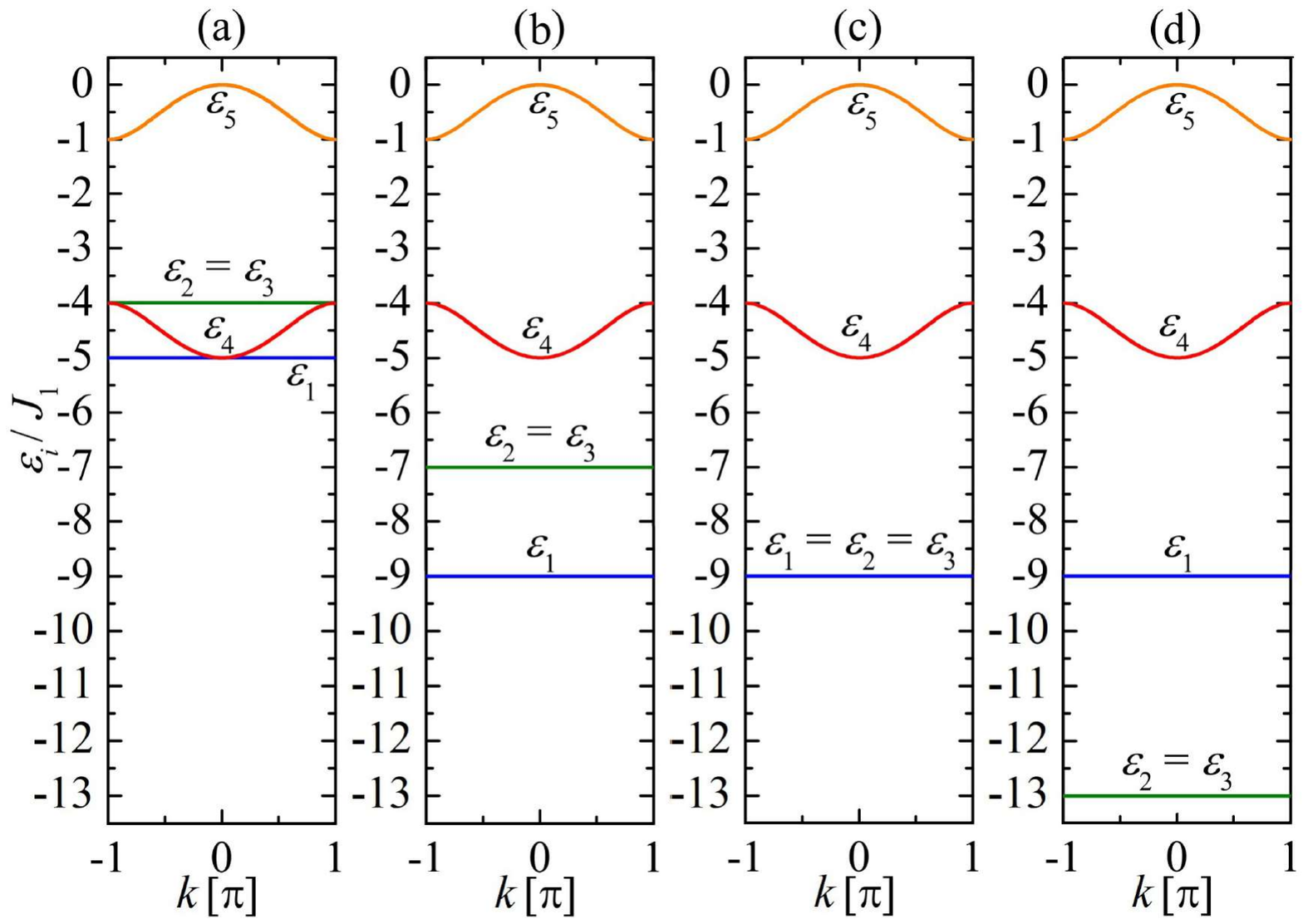}
\end{center}
\vspace{-0.6cm}
\caption{(Color online) The one-magnon energy bands (\ref{oms}) of the spin-$\frac{1}{2}$ Heisenberg octahedral chain for four representative values of the interaction ratio: 
(a) $J_2/J_1 = 2$, $J_3/J_1 = 1$; (b) $J_2/J_1 = 4$, $J_3/J_1 = 2$; (c) $J_2/J_1 = 4$, $J_3/J_1 = 4$; (d) $J_2/J_1 = 4$, $J_3/J_1 = 8$.}
\label{fig2}
\end{figure}

In particular, the flat band with the energy cost $\varepsilon_{1}=- J_1 - 2J_2$ has the lowest energy in the highly frustrated regime with dominant nearest-neighbour coupling constant within the square plaquettes $J_2 > 2 J_1$ and $J_2 > J_3$. This flat band corresponds to a bound one-magnon state supporting a single localized magnon trapped within a square plaquette:
\begin{eqnarray}
|1,1,1 \rangle_j \!&=&\! |S_{\square,j}=1, S_{24,j}= 1, S_{35,j}=1 \rangle_j \nonumber \\
\!&=&\! \frac{1}{2}
(|\!\!\downarrow_{2,j}\uparrow_{3,j}\uparrow_{4,j}\uparrow_{5,j}\rangle 
\!-\!|\!\!\uparrow_{2,j}\downarrow_{3,j}\uparrow_{4,j}\uparrow_{5,j}\rangle \nonumber \\
\!&+&\! |\!\!\uparrow_{2,j}\uparrow_{3,j}\downarrow_{4,j}\uparrow_{5,j}\rangle
\!-\!|\!\!\uparrow_{2,j}\uparrow_{3,j}\uparrow_{4,j}\downarrow_{5,j}\rangle). 
\label{om1}
\end{eqnarray}
The many-magnon crystal states of the spin-$\frac{1}{2}$ Heisenberg octahedral chain can be systematically constructed by populating the square plaquettes with independent localized magnons of the form (\ref{om1}). It can be easily verified that the lowest-energy state in the parameter regime $J_2 > 2 J_1$, $J_2 > J_3$ and $h<J_1 + 2 J_2$ is the bound magnon crystal with the highest possible number of localized one-magnon plaquette (OMP) states: 
\begin{eqnarray}
|{\rm OMP}\rangle \!=\! \prod_{j=1}^N \! |\!\!\uparrow_{1,j}\rangle \!\otimes\! \frac{1}{2}
(|\!\!\downarrow_{2,j}\uparrow_{3,j}\uparrow_{4,j}\uparrow_{5,j}\rangle 
\!\!&-&\!\!|\!\!\uparrow_{2,j}\downarrow_{3,j}\uparrow_{4,j}\uparrow_{5,j}\rangle \nonumber \\
+|\!\!\uparrow_{2,j}\uparrow_{3,j}\downarrow_{4,j}\uparrow_{5,j}\rangle
\!\!&-&\!\!|\!\!\uparrow_{2,j}\uparrow_{3,j}\uparrow_{4,j}\downarrow_{5,j}\rangle). \nonumber \\  
\label{OMP}
\end{eqnarray}
The bound magnon-crystal phase (\ref{OMP}) gives rise to an intermediate three-fifths plateau in the zero-temperature magnetization curves, which is limited to the field range $h \in (J_1 + J_2, J_1 + 2J_2)$ as lower magnetic fields favor the fragmented TMP crystal (\ref{TMP}).

The highly frustrated regime with the dominant next-nearest-neighbour interaction along the diagonals of a square plaquette $J_3 > 2 J_1$ and $J_3 > J_2$ contrarily favors a two-fold degenerate flat band with the energy cost $\varepsilon_{2,3}=- J_1 - J_2 - J_3$, which corresponds to the formation of a dimer-singlet along one of the two diagonals of a square plaquette:  
\begin{eqnarray}
|1,0,1 \rangle_j \!&=&\! |S_{\square,j}=1, S_{24,j}= 0, S_{35,j}=1 \rangle_j \nonumber \\
\!&=&\! \frac{1}{\sqrt{2}} (|\!\!\uparrow_{2,j}\downarrow_{4,j}\rangle \!-\!|\!\!\downarrow_{2,j}\uparrow_{4,j}\rangle) \!\otimes\! |\!\! \uparrow_{3,j} \uparrow_{5,j} \rangle, \nonumber \\
|1,1,0 \rangle_j \!&=&\! |S_{\square,j}=1, S_{24,j}= 1, S_{35,j}=0 \rangle_j \nonumber \\
\!&=&\! |\!\! \uparrow_{2,j} \uparrow_{4,j} \rangle \!\otimes\! \frac{1}{\sqrt{2}} (|\!\!\uparrow_{3,j}\downarrow_{5,j}\rangle \!-\!|\!\!\downarrow_{3,j}\uparrow_{5,j}\rangle). 
\label{om23}
\end{eqnarray}
The bound one-magnon states  (\ref{om23}) can be repeatedly used to construct a series of many-magnon crystals of the spin-$\frac{1}{2}$ Heisenberg octahedral chain by populating the square plaquettes with independent localized magnons of this form. It can be proved that the bound magnon crystal with the highest possible number of localized one-magnon dimer-singlet (OMD) states represents the respective ground state of the spin-$\frac{1}{2}$ Heisenberg octahedral chain in the parameter region $J_3 > 2 J_1$, $J_3 > J_2$ and $h<J_1 + J_2 + J_3$: 
\begin{eqnarray}
|{\rm OMD}\rangle \!\!=\!\! \prod_{j=1}^N \! |\!\!\uparrow_{1,j}\rangle \!\otimes\! 
\biggl\{ \!\! \begin{array}{l} 
\! \frac{1}{\sqrt{2}} (|\!\!\uparrow_{2,j}\downarrow_{4,j}\rangle \!-\!|\!\!\downarrow_{2,j}\uparrow_{4,j}\rangle) \!\otimes\! |\!\! \uparrow_{3,j} \uparrow_{5,j} \rangle \\
\! |\!\! \uparrow_{2,j} \uparrow_{4,j} \rangle \!\otimes\! \frac{1}{\sqrt{2}} (|\!\!\uparrow_{3,j}\downarrow_{5,j}\rangle \!-\!|\!\!\downarrow_{3,j}\uparrow_{5,j}\rangle)
\end{array}\!\!, \biggr. \nonumber \\
\label{OMD}
\end{eqnarray}
The bound magnon-crystal phase (\ref{OMD}) has a macroscopic degeneracy $2^N$ as each square plaquette can host two distinct dimer singlets on one of its diagonals. Similarly to the previous case,  the magnon crystal (\ref{OMD}) gives rise to an intermediate three-fifths plateau in the zero-temperature magnetization curves, which is limited to the field range $h \in (J_1 + J_3, J_1 + J_2 + J_3)$ as lower magnetic fields favor the fragmented TMP crystal (\ref{TMP}).

\subsection{Extended localized-magnon theory} 
\label{lmt}

In the highly frustrated regime where at least one of the intra-plaquette coupling constants exceeds twice the strength of the monomer-plaquette coupling constant, i.e. $J_2/J_1 > 2$ or $J_3/J_1 > 2$, we elaborate an extended localized-magnon theory to elucidate the low-temperature features of the spin-$\frac{1}{2}$ Heisenberg octahedral chain. Several exact many-magnon eigenstates of the model can be systematically constructed from five localized-magnon states: two distinct bound two-magnon states corresponding to the plaquette-singlet state (\ref{PS}) and the direct product of two dimer-singlet states (\ref{DS}), as well as, three bound one-magnon states associated with the collective plaquette state (\ref{om1}) and dimer-singlet states (\ref{om23}) formed along one of two diagonals of the square plaquette. Each square plaquette can independently host one of these localized-magnon states and hence, the low-temperature features of the spin-$\frac{1}{2}$ Heisenberg octahedral chain can be effectively mapped onto a five-component lattice-gas model of hard-core monomers, where each monomer species represents a distinct type of bound magnon state. 

Within this effective lattice-gas framework, the bound one-magnon plaquette state (\ref{om1}) is represented by the first type  of particle with the chemical potential $\mu_1^{(1)} = J_1 + 2J_2 - h$ and the corresponding occupation number $n_{1,j}^{(1)} = 0,1$, while the other two bound one-magnon states (\ref{om23}) with the character of dimer singlets are represented by the second and third types of particle both having identical chemical potentials $\mu_1^{(2,3)} = J_1 + J_2 + J_3 - h$ and the respective occupation numbers $n_{1,j}^{(2)} = 0,1$ and $n_{1,j}^{(3)} = 0,1$. Similarly, the bound two-magnon plaquette state (\ref{PS}) is represented by the fourth type of particle with the chemical potential $\mu_2^{(1)} = J_1 + 3J_2/2 - 2h$ and the occupation number $n_{2,j}^{(1)} = 0,1$, whereas the two-magnon product state of two dimer singlets (\ref{DS}) is represented by the fifth particle type with the chemical potential $\mu_2^{(2)} = J_1 + J_2/2 + J_3 - 2h$ and the corresponding occupation number $n_{2,j}^{(2)} = 0,1$. All chemical potentials are defined so as to reflect the energy cost required to create a particular bound magnon state on top of the fully polarized ferromagnetic background serving as a vacuum state. The Hamiltonian of the resulting effective lattice-gas model can thus be written as:
\begin{eqnarray}
{\cal H}_{\rm eff} = E_{\rm FM}^0 \!&-&\! h\left(2N + \sum_{j=1}^{N} S_{1,j}^z\right) - \mu_1^{(1)} \sum_{j=1}^{N} n_{1,j}^{(1)} \nonumber \\
\!&-&\! \mu_1^{(2,3)} \left(\sum_{j=1}^{N} n_{1,j}^{(2)}+\sum_{j=1}^{N} n_{1,j}^{(3)}\right) \nonumber \\
\!&-&\! \mu_2^{(1)} \sum_{j=1}^{N} n_{2,j}^{(1)} - \mu_2^{(2)} \sum_{j=1}^{N} n_{2,j}^{(2)}. 
\label{elm}
\end{eqnarray} 

The Zeeman term $-h(2N + \sum_{j=1}^{N} S_{1,j}^z)$ specifically involves separate contributions from the plaquette and monomer spins, respectively. The variable contribution of the monomer spins to the Zeeman energy reflects their possible paramagnetic character whenever both square plaquettes adjacent to a given monomer spin are either in a bound two-magnon state with the character of the plaquette singlet (\ref{PS}) or the product of two diagonal dimer singlets (\ref{DS}). The partition function of the effective lattice-gas model accounting for all available bound many-magnon states of the spin-$\frac{1}{2}$ Heisenberg octahedral chain then follows from the formula:
\begin{widetext}
\begin{eqnarray}
{\cal Z} &=& \exp \left(-\beta E_{\rm FM}^0 + 2\beta Nh \right) \!\! \sum_{\{S_{1,j}^z\}} \prod_{j=1}^N \biggl\{
 \biggl[\sum_{\alpha, \beta, \gamma, \delta} \sum_{n_{\alpha,j}^{(\beta)}} \sum_{n_{\gamma,j}^{(\delta)}} 
(1 - n_{\alpha,j}^{(\beta)} n_{\gamma,j}^{(\delta)})\biggr] \exp \left[\frac{\beta h}{2} \left(S_{1,j}^z + S_{1,j+1}^z\right)  \right]  
\label{pfef} \\
 &\times&  \sum_{n_{1,j}^{(1)}} \sum_{n_{1,j}^{(2)}} \sum_{n_{1,j}^{(3)}} \sum_{n_{2,j}^{(1)}} \sum_{n_{2,j}^{(2)}} 
\left[\left( \frac{1}{2} \!+\! S_{1,j}^z \!\right) \! \left(\frac{1}{2} \!+\! S_{1,j+1}^z \!\right) 
\exp \left[\beta \mu_1^{(1)} n_{1,j}^{(1)} \!+\! \beta \mu_1^{(2,3)} (n_{1,j}^{(2)} \!+\! n_{1,j}^{(3)}) \right] \!\!
+ \exp \left(\beta \mu_2^{(1)} n_{2,j}^{(1)} \!+\! \beta \mu_2^{(2)} n_{2,j}^{(2)} \right)\right] \biggr\}\!,  \nonumber 
\end{eqnarray}
\end{widetext}
where $\beta = 1/(k_{\rm B} T)$, $k_{\rm B}$ is Boltzmann's constant, $T$ is the absolute temperature, and the product of projection operators $\sum_{\alpha, \beta, \gamma, \delta} \sum_{n_{\alpha,j}^{(\beta)}} \sum_{n_{\gamma,j}^{(\delta)}} (1 - n_{\alpha,j}^{(\beta)} n_{\gamma,j}^{(\delta)})$ prevents each square plaquette from hosting more than one bound magnon state (particle). After performing the summation over all occupation numbers the partition function can be expressed in the following form: 
\begin{eqnarray}
{\cal Z} = {\rm e}^{-\beta E_{\rm FM}^0 + 2\beta Nh} \sum_{\{S_{1,j}^z\}} \prod_{j=1}^N T(S_{1,j}^z ; S_{1,j+1}^z),
\label{pftm}
\end{eqnarray}
which involves the Boltzmann weight $T(S_{1,j}^z ; S_{1,j+1}^z)$ depending on two adjacent monomer spins identified as the transfer matrix:
\begin{eqnarray}
&T&(S_{1,j}^z ; S_{1,j+1}^z) = \exp \left[\frac{\beta h}{2} \left(S_{1,j}^z + S_{1,j+1}^z\right) \right] \nonumber \\
&& \left[ \left( \frac{1}{2} + S_{1,j}^z\right) \left(\frac{1}{2} + S_{1,j+1}^z \right) \left(1 + w_1 \right) 
+ w_2 \right]\!. 
\label{tm}
\end{eqnarray}
Here, the expressions $w_1 = {\rm e}^{\beta\mu_1^{(1)}} + 2{\rm e}^{\beta\mu_1^{(2,3)}}$ and $w_2 = {\rm e}^{\beta\mu_2^{(1)}} + {\rm e}^{\beta\mu_2^{(2)}}$ denote the total Boltzmann weights of all bound one- and two-magnon states, respectively. Performing the consecutive summation over spin states of the monomer spins within the standard transfer-matrix approach yields the following expression for the partition function:
\begin{eqnarray}
{\cal Z} = {\rm e}^{-\beta E_{\rm FM}^0 + 2\beta Nh} {\rm Tr} \, T^N \!\!
= {\rm e}^{-\beta E_{\rm FM}^0 + 2\beta Nh} \! \left(\lambda_+^N \!+\! \lambda_-^N\right)\!,
\label{pftme}
\end{eqnarray}
which is expressed in terms of the transfer-matrix eigenvalues:
\begin{eqnarray}
\lambda_{\pm} &=& \frac{1}{2} \biggl\{ {\rm e}^{\frac{\beta h}{2}} 
\left( 1 + w_1 + w_2 \right) + {\rm e}^{-\frac{\beta h}{2}} w_2  \nonumber \\
&\pm& \sqrt{\left[ {\rm e}^{\frac{\beta h}{2}} \left( 1 + w_1 + w_2 \right) - {\rm e}^{-\frac{\beta h}{2}} w_2 \right]^2 
\!\! + 4 w_2^2} \biggr\}\!.
\label{lmmag}
\end{eqnarray}
In the thermodynamic limit $N \to \infty$, the density of free energy normalized per spin can be calculated from the relation:
\begin{eqnarray}
f &=& - k_{\rm B} T \lim_{N \to \infty} \frac{1}{5N} \ln {\cal Z} \nonumber \\
  &=& \frac{1}{10} (4J_1 + 2J_2 + J_3) - \frac{2}{5}h - \frac{k_{\rm B} T}{5} \ln \lambda_+.
\label{lmgfe}
\end{eqnarray}
The free-energy density (\ref{lmgfe}) can be subsequently utilized for a calculation of the magnetization, susceptibility, entropy, and specific heat per spin according to the standard relations:
\begin{eqnarray}
m = -\frac{\partial f}{\partial h}, \, \chi = -\frac{\partial^2 f}{\partial h^2}, \, 
s = -\frac{\partial f}{\partial T}, \, c = -T \frac{\partial^2 f}{\partial T^2}.
\label{mssc}
\end{eqnarray}

\subsection{Exact diagonalization} 

To verify the validity of the extended localized-magnon theory, we have also carried out the full exact diagonalization (ED) calculations for a finite-size spin-$\frac{1}{2}$ Heisenberg octahedral chain containing up to $20$ spins ($N=4$ unit cells). For this purpose, we adapted the subroutine fulldiag from the open-source Algorithms and Libraries for Physics Simulation (ALPS) package \cite{baue11}, which enables an exact numerical diagonalization of the full Hamiltonian matrix and thus provides a reliable benchmark for testing analytical predictions.

\begin{figure*}
\begin{center}
\includegraphics[width=0.5\textwidth]{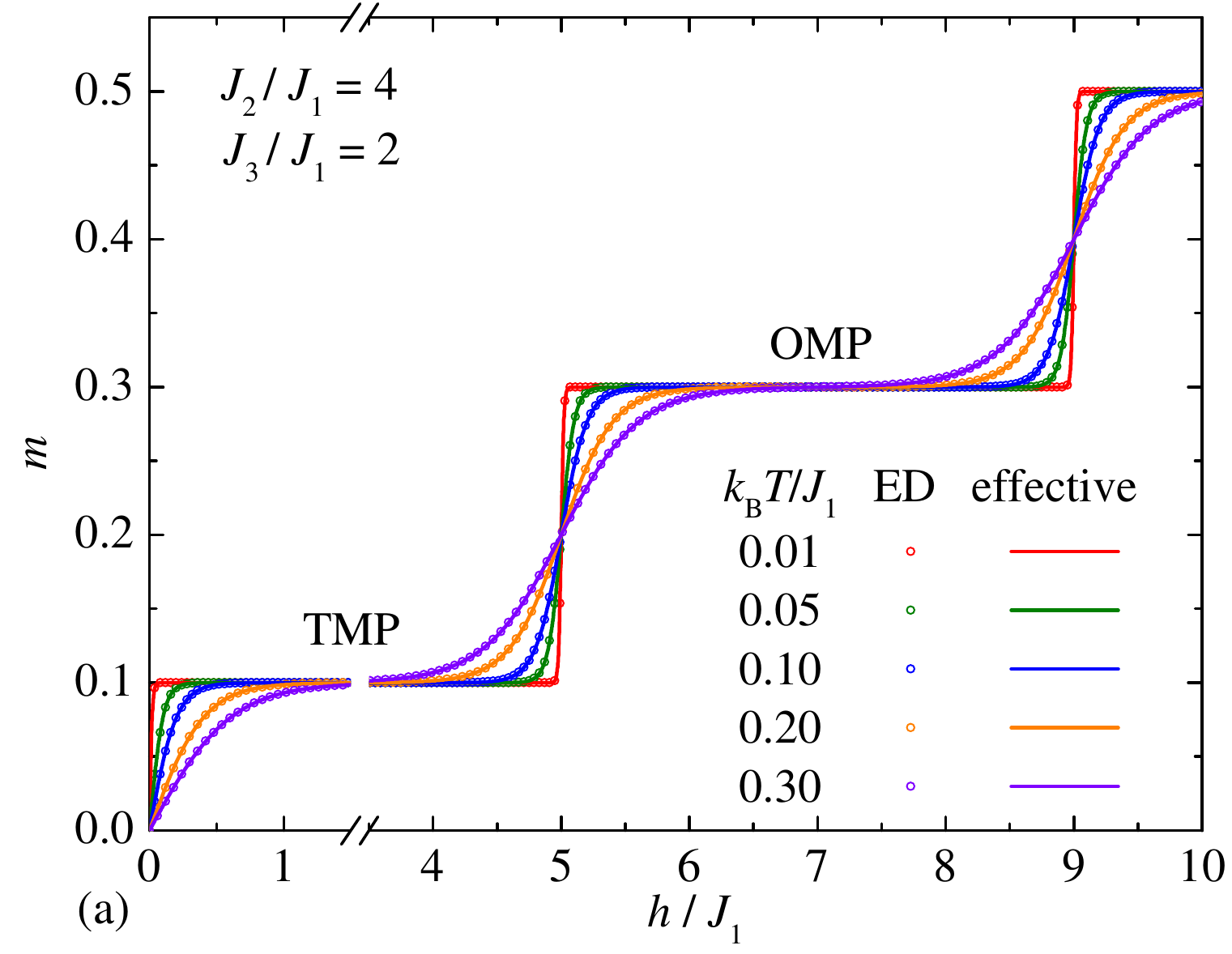}
\hspace{-0.2cm}
\includegraphics[width=0.5\textwidth]{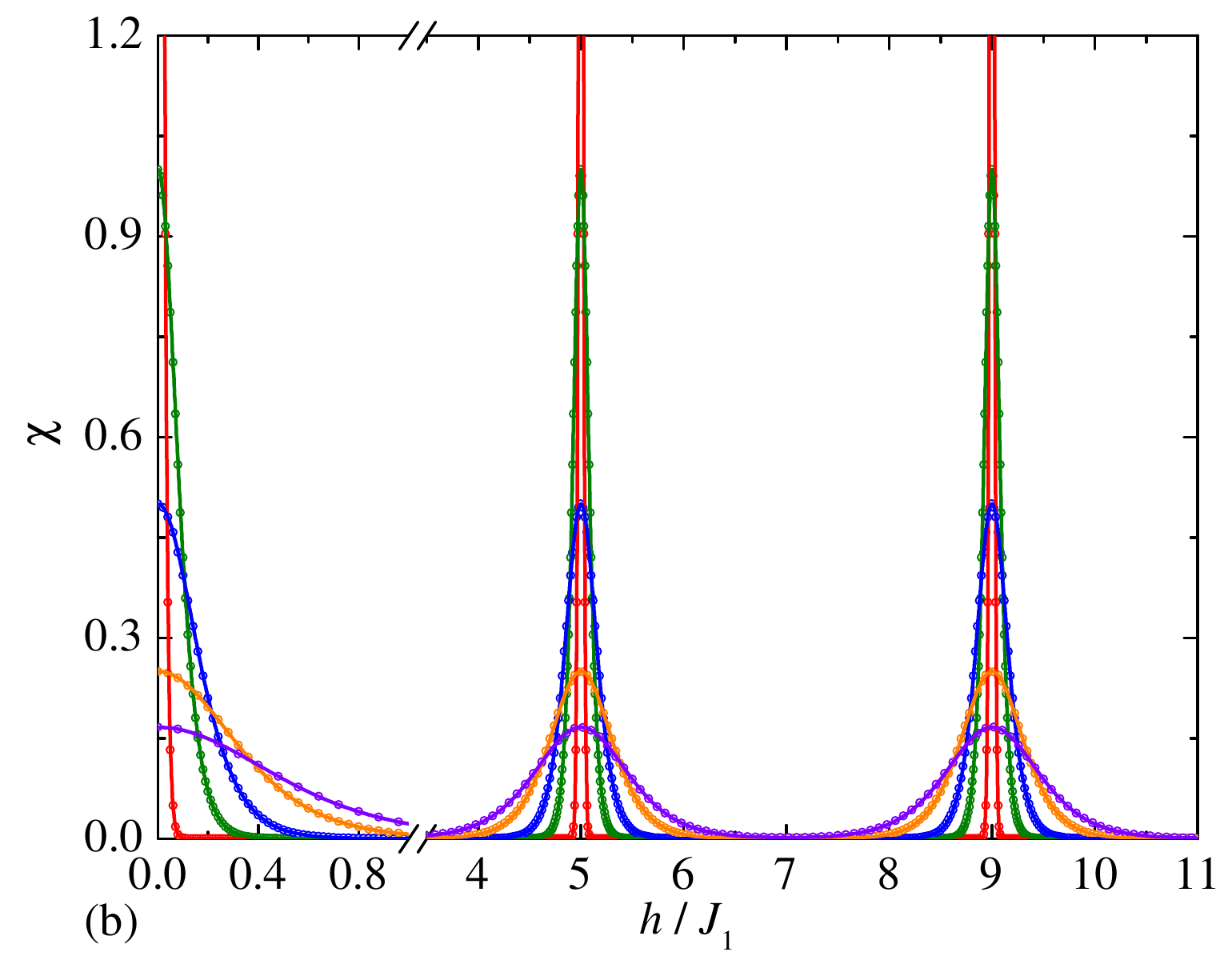}
\includegraphics[width=0.5\textwidth]{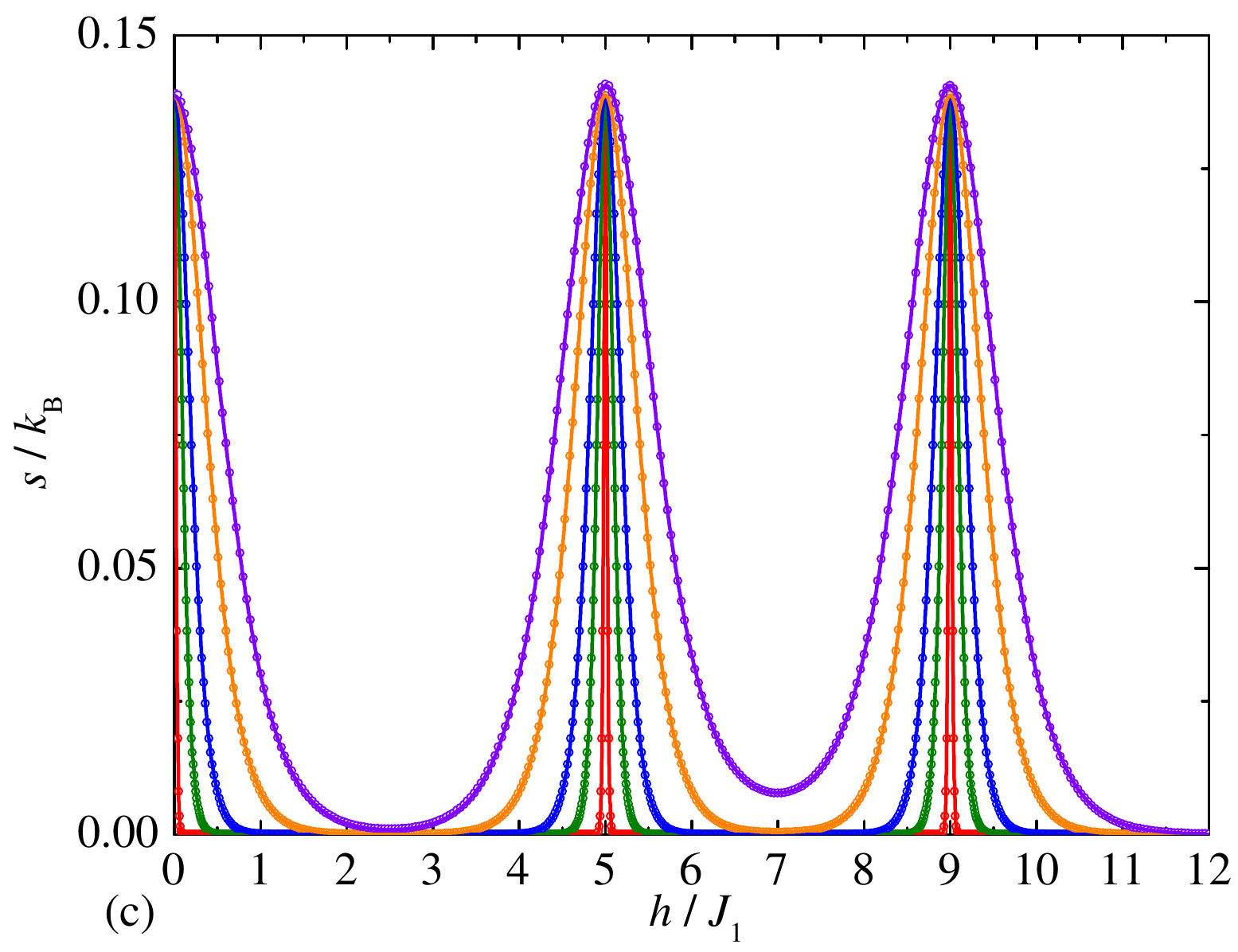}
\hspace{-0.2cm}
\includegraphics[width=0.5\textwidth]{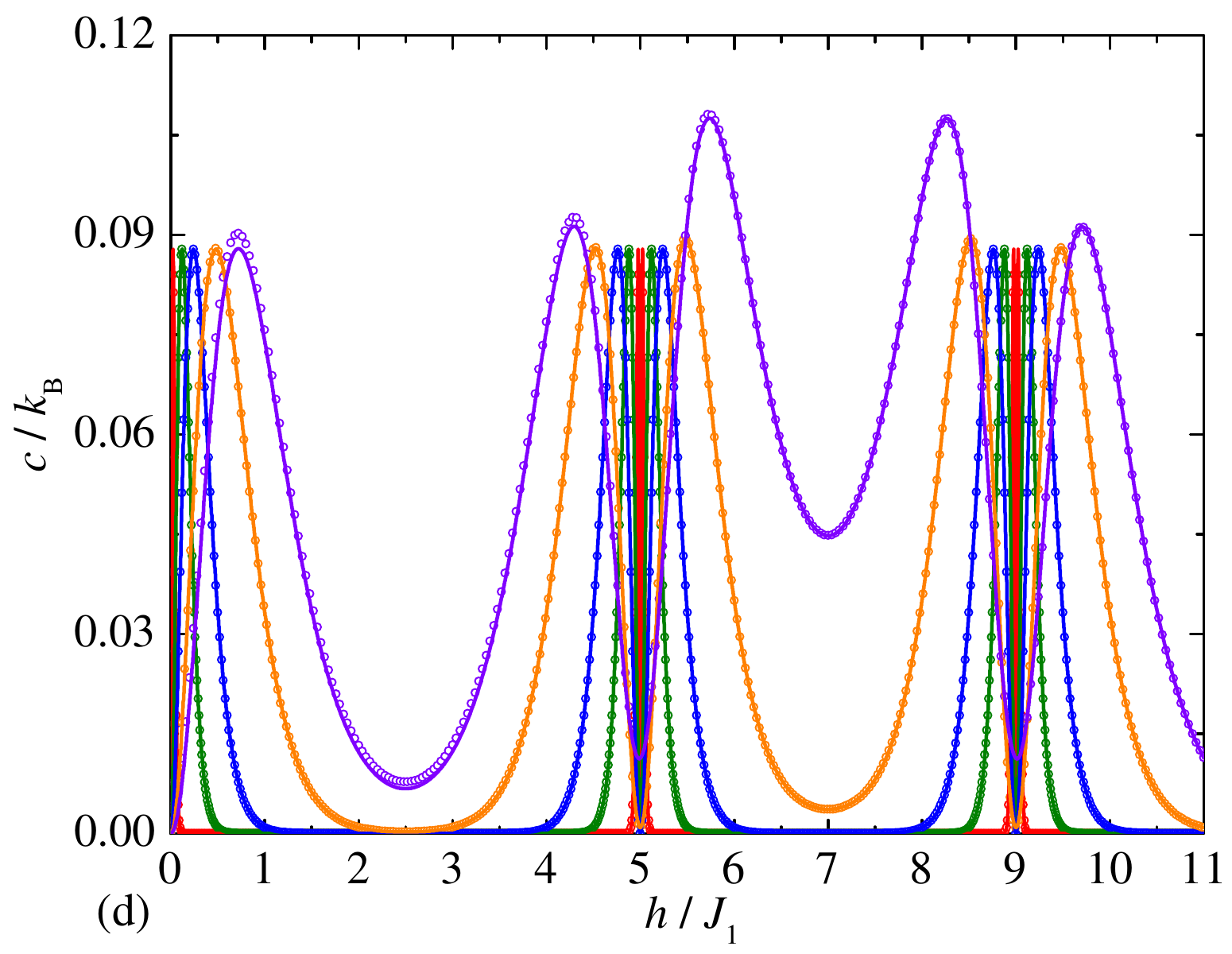}
\end{center}
\vspace{-0.8cm}
\caption{A few isothermal field variations of the magnetization (a), susceptibility (b), entropy (c), and specific heat (d) of the spin-$\frac{1}{2}$ Heisenberg octahedral chain with the interaction ratio $J_2/J_1 = 4$ and $J_3/J_1 = 2$. Symbols show ED data for the finite chain of 20 spins, while solid lines were derived from the analytical expression (\ref{lmgfe}) for the free energy of the effective lattice-gas model in the thermodynamic limit. The legend in the panel (a) applies also to panels (b)-(d).}
\label{fig3}
\end{figure*}

\section{Results and discussion}
\label{sec:result}

In this section, we present a comprehensive analysis of the most significant low-temperature magnetic and thermodynamic properties of the spin-$\frac{1}{2}$ Heisenberg octahedral chain for three representative parameter regimes supporting various bound magnon crystal states.  

\begin{figure*}
\begin{center}
\includegraphics[width=0.5\textwidth]{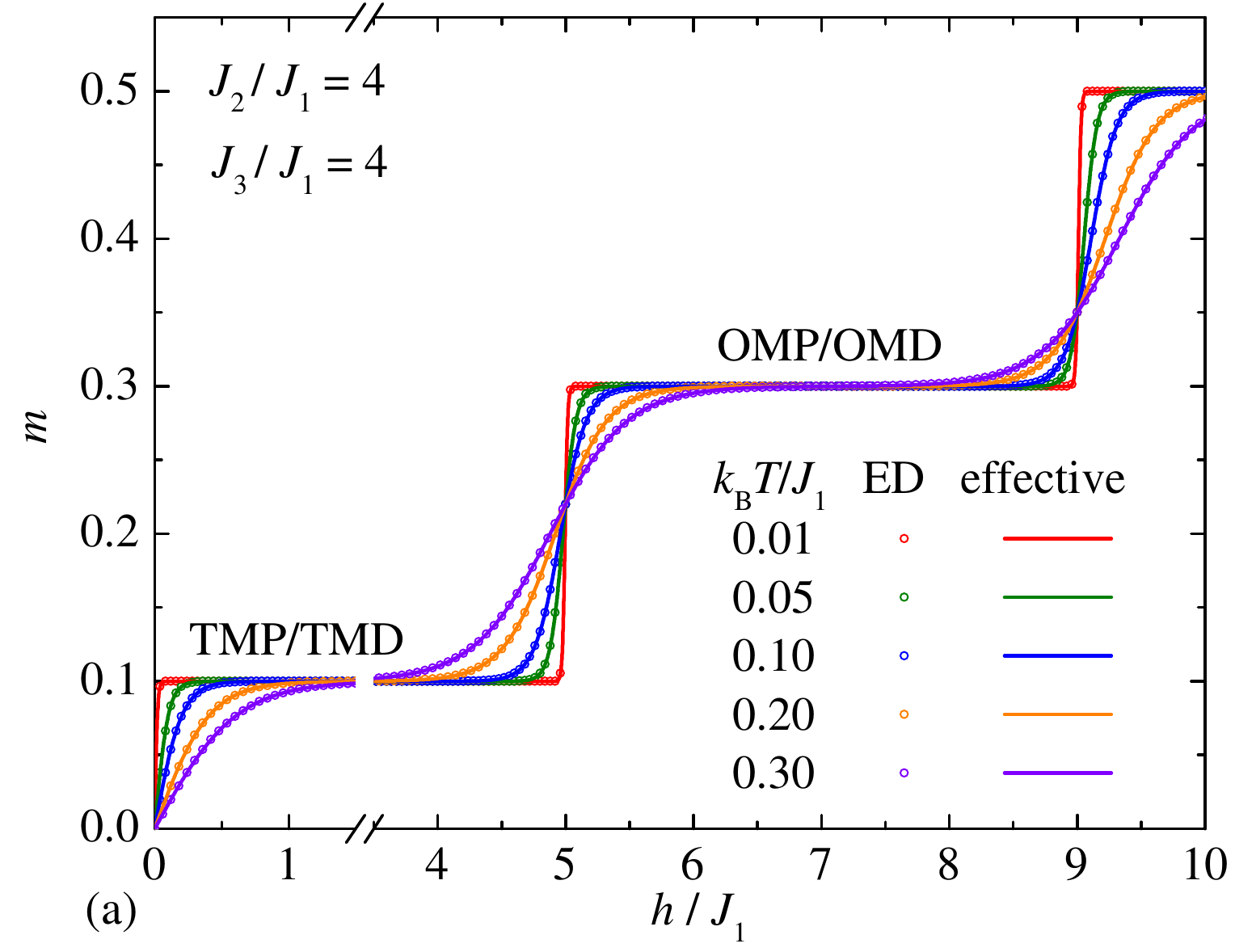}
\hspace{-0.2cm}
\includegraphics[width=0.5\textwidth]{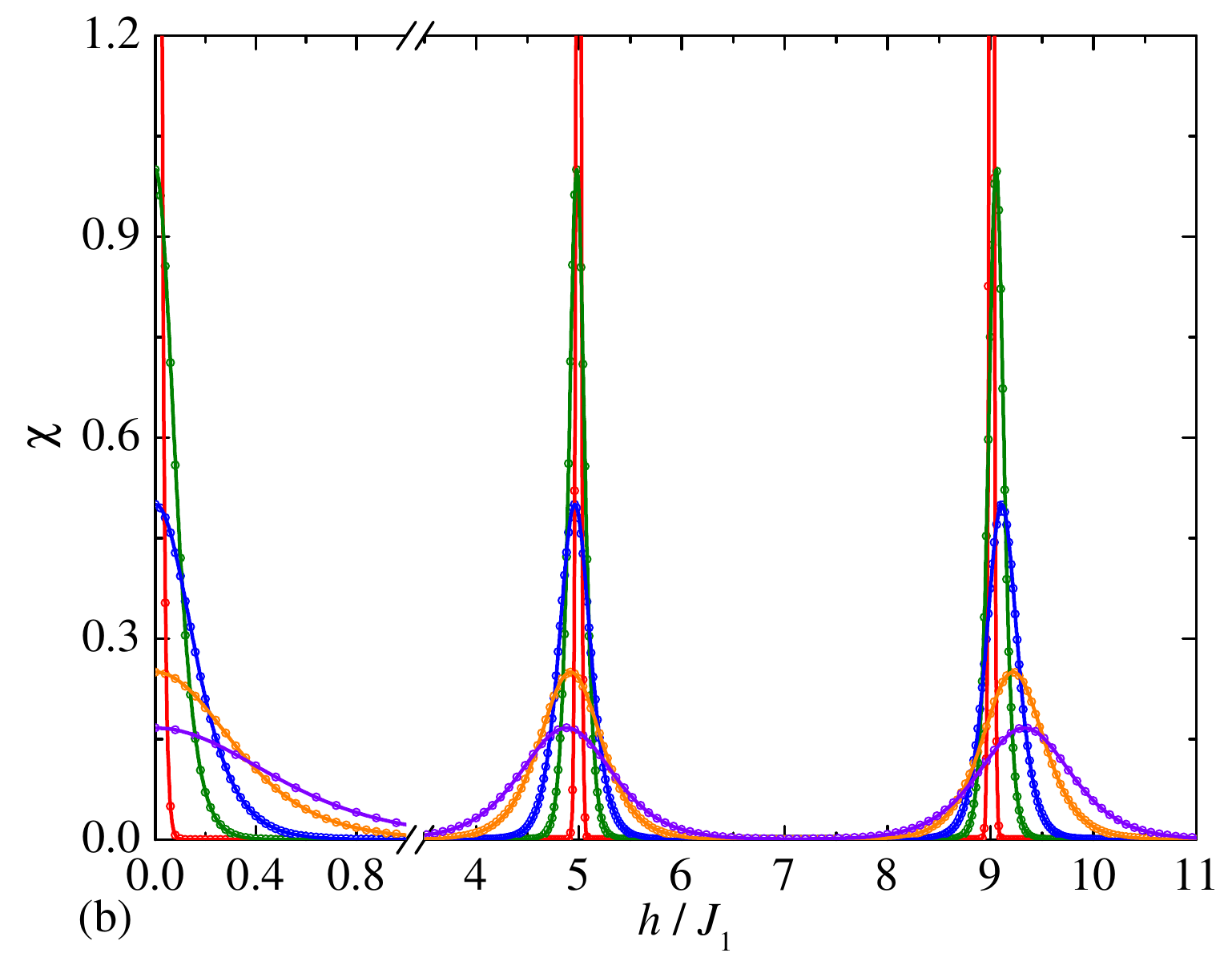}
\includegraphics[width=0.5\textwidth]{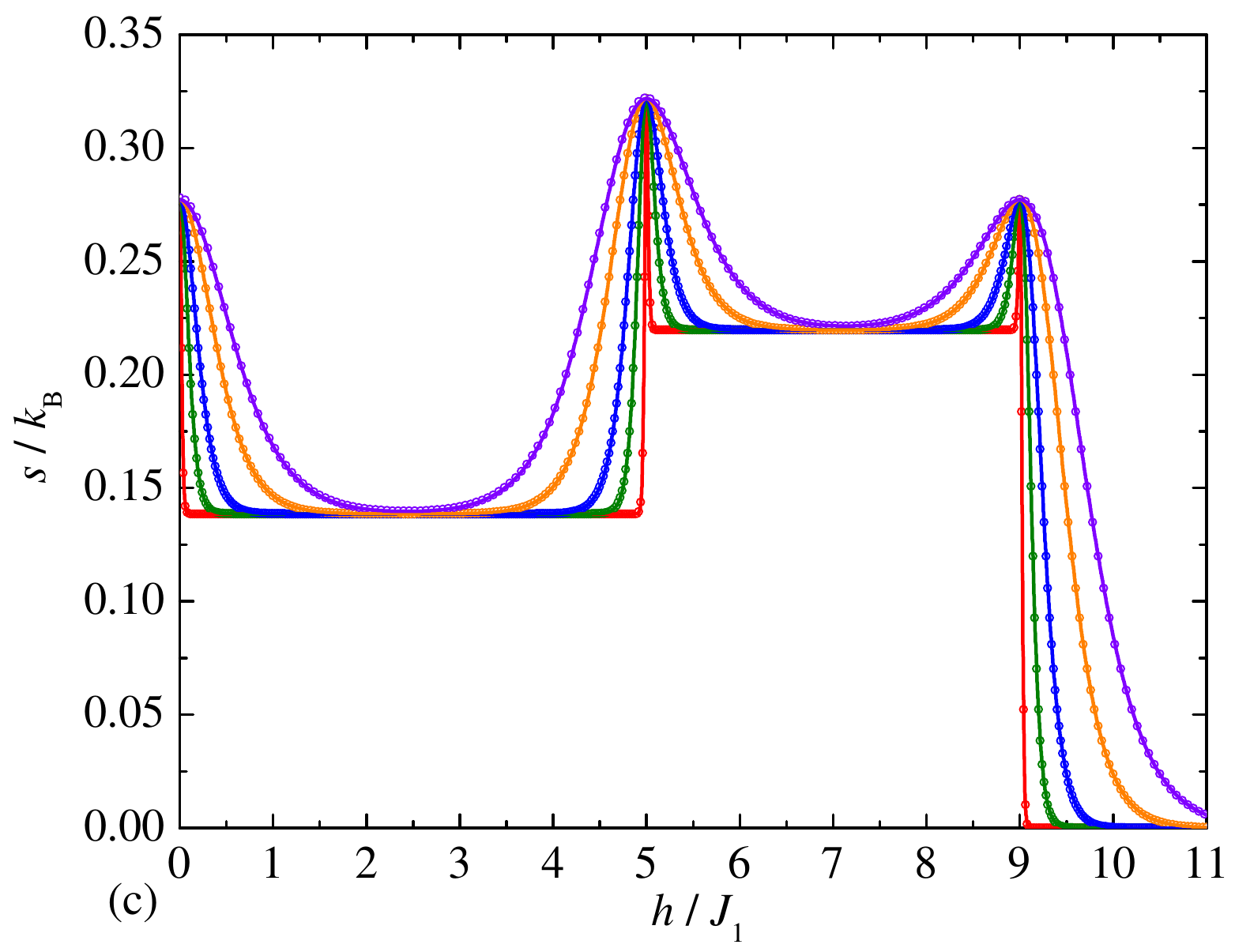}
\hspace{-0.2cm}
\includegraphics[width=0.5\textwidth]{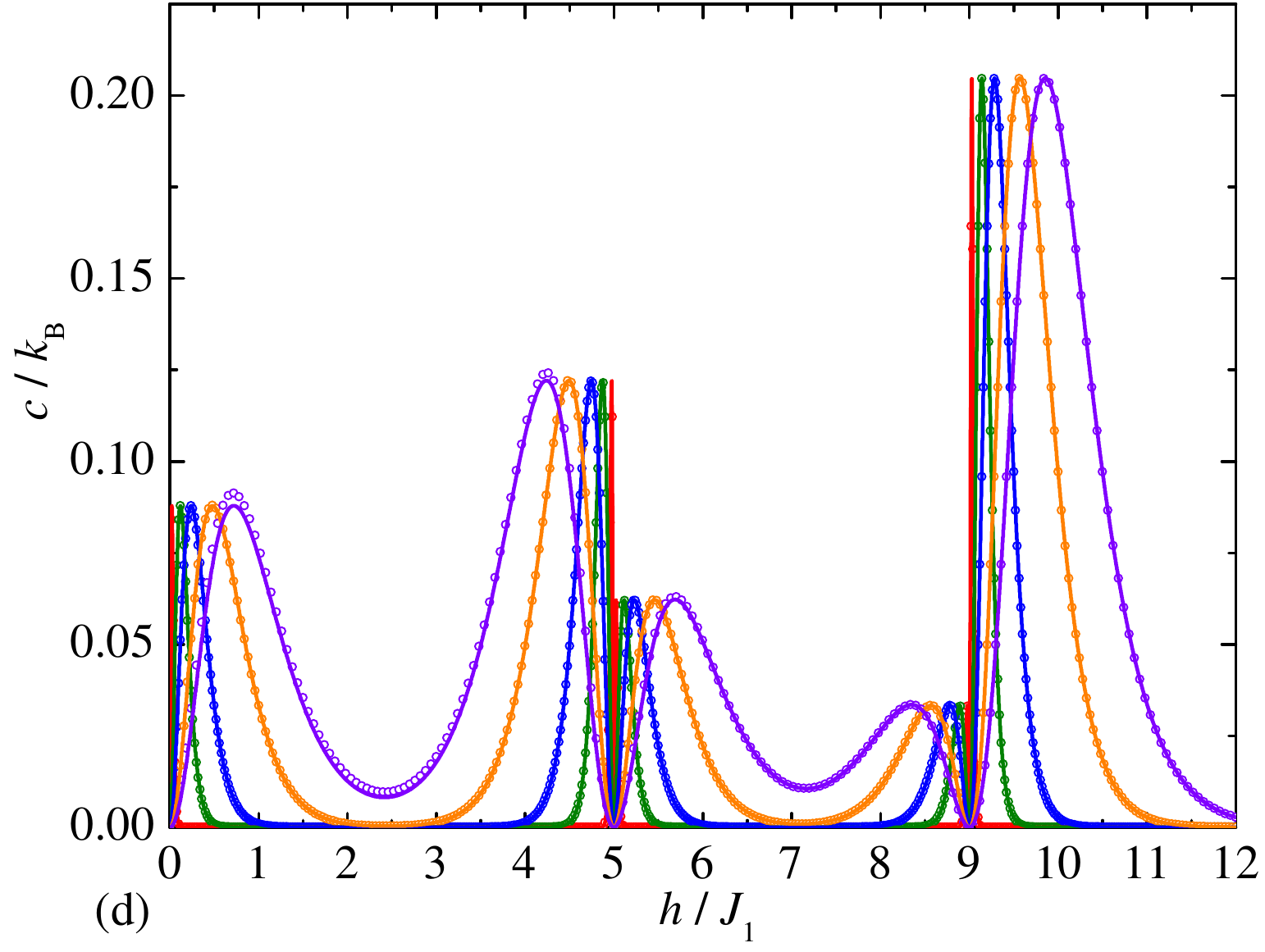}
\end{center}
\vspace{-0.8cm}
\caption{A few isothermal field variations of the magnetization (a), susceptibility (b), entropy (c), and specific heat (d) of the spin-$\frac{1}{2}$ Heisenberg octahedral chain with the interaction ratio $J_2/J_1 = 4$ and $J_3/J_1 = 4$. Symbols show ED data for the finite chain of 20 spins, while solid lines were derived from the analytical expression (\ref{lmgfe}) for the free energy of the effective lattice-gas model in the thermodynamic limit. The legend in the panel (a) applies also to panels (b)-(d).}
\label{fig4}
\end{figure*} 

\subsection{Collective bound-magnon plaquette states}

First, we examine the spin-$\frac{1}{2}$ Heisenberg octahedral chain with fixed values of the interaction ratio $J_2/J_1 = 4$ and $J_3/J_1 = 2$, which supports at sufficiently low magnetic fields bound two- and one-magnon plaquette phases TMP and OMP described by Eqs. (\ref{PS}) and (\ref{OMP}), respectively. A few typical isothermal field variations of the magnetization, magnetic susceptibility, magnetic entropy, and magnetic specific heat are plotted in Fig. \ref{fig3} for this parameter set. The numerical results obtained from ED for a finite chain of 20 spins (symbols) are compared with the respective theoretical predictions based on the five-component lattice-gas model in the thermodynamic limit (solid lines). It is obvious from Fig. \ref{fig3} that two data sets exhibit nearly perfect quantitative agreement up to moderate temperatures $k_{\rm B} T/J_1 \approx 0.3$ which confirms negligible finite-size effects and validates the lattice-gas description including microscopic nature of the underlying magnon-crystal phases. Two-step magnetization curves with intermediate plateaus at one-fifth and three-fifths of the saturation magnetization separated by abrupt magnetization changes occurring near transition fields $h/J_1 = 5$ and $9$ are indeed fully consistent with successive field-driven transitions involving the bound magnon-crystal phases TMP and OMP [see Fig. \ref{fig3}(a)]. Increasing temperature gradually smears out a stepwise profile of the magnetization curves, which intersect each other at a temperature-independent crossing point located at the midpoint of each magnetization jump. This thermal effect is also manifested by the broadening and suppression of the susceptibility peaks centered at the transition fields as illustrated in Fig. \ref{fig3}(b). A similar broadening trend is observed in the field dependence of the entropy shown in Fig. \ref{fig3}(c) though the peak height remains nearly constant at $s/k_{\rm B} = \frac{1}{5} \ln 2 \approx 0.1386$. This specific value reflects the macroscopic degeneracy $2^N$ of the ground state at the transition field arising from a two-fold degeneracy of each square plaquette. Finally, the specific heat displays a pronounced double-peak structure symmetrically centered around each transition field in Fig. \ref{fig3}(d). The observed peak height $c/k_{\rm B} \approx 0.0878$ is consistent with the Schottky maximum of a two-level system allowing their interpretation as thermally activated excitations between two unique non-degenerate states.     

\begin{figure*}
\begin{center}
\includegraphics[width=0.5\textwidth]{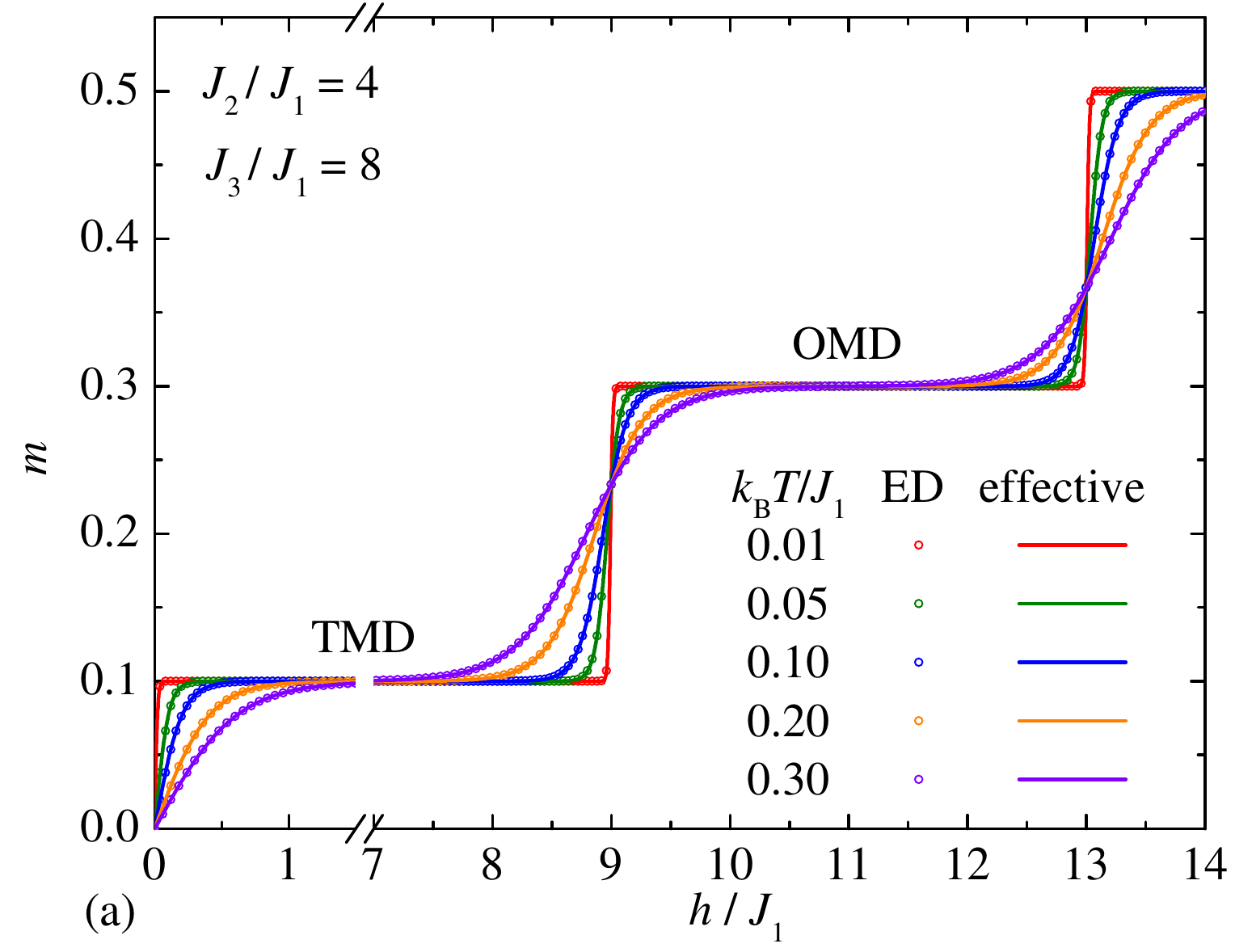}
\hspace{-0.2cm}
\includegraphics[width=0.5\textwidth]{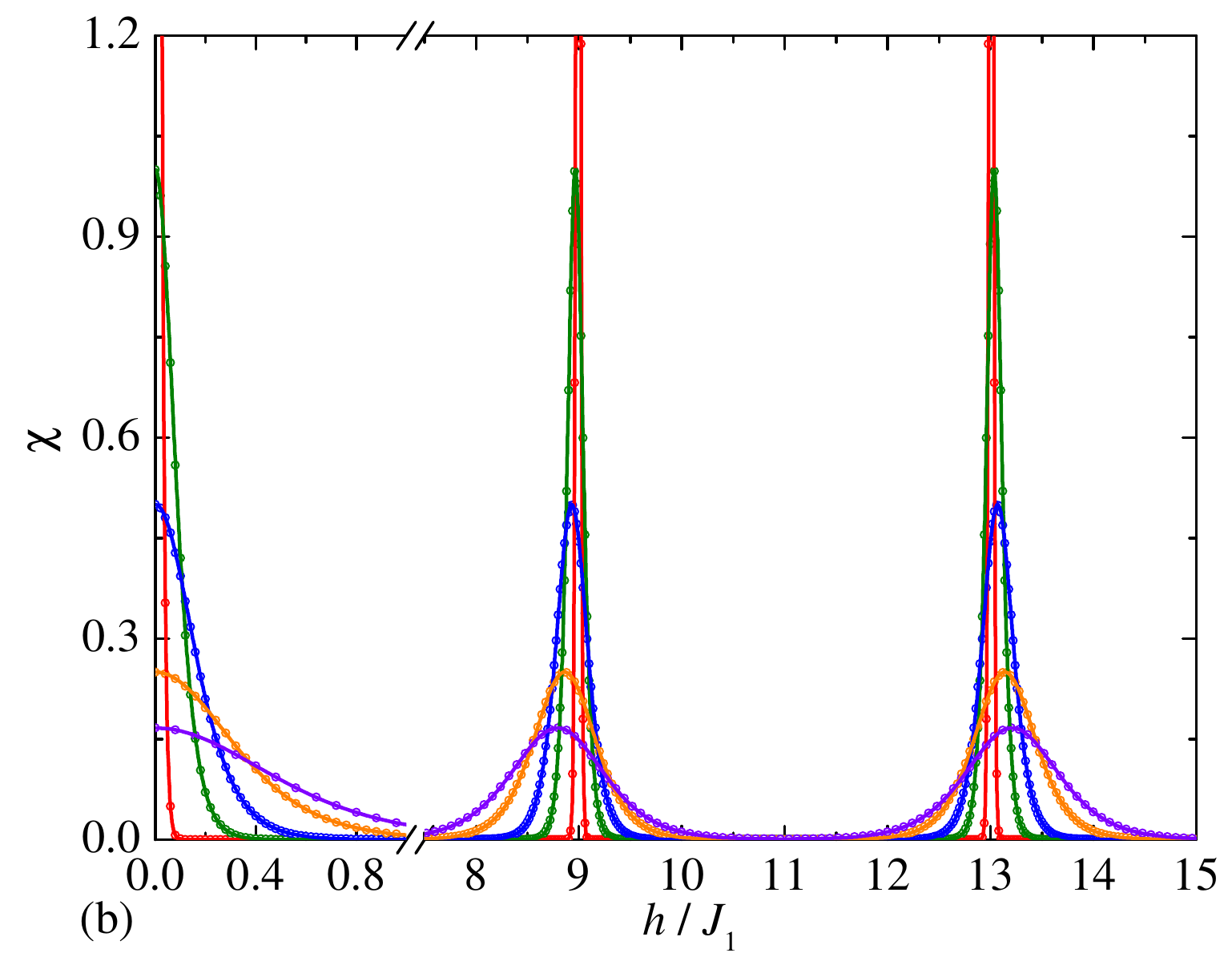}
\includegraphics[width=0.5\textwidth]{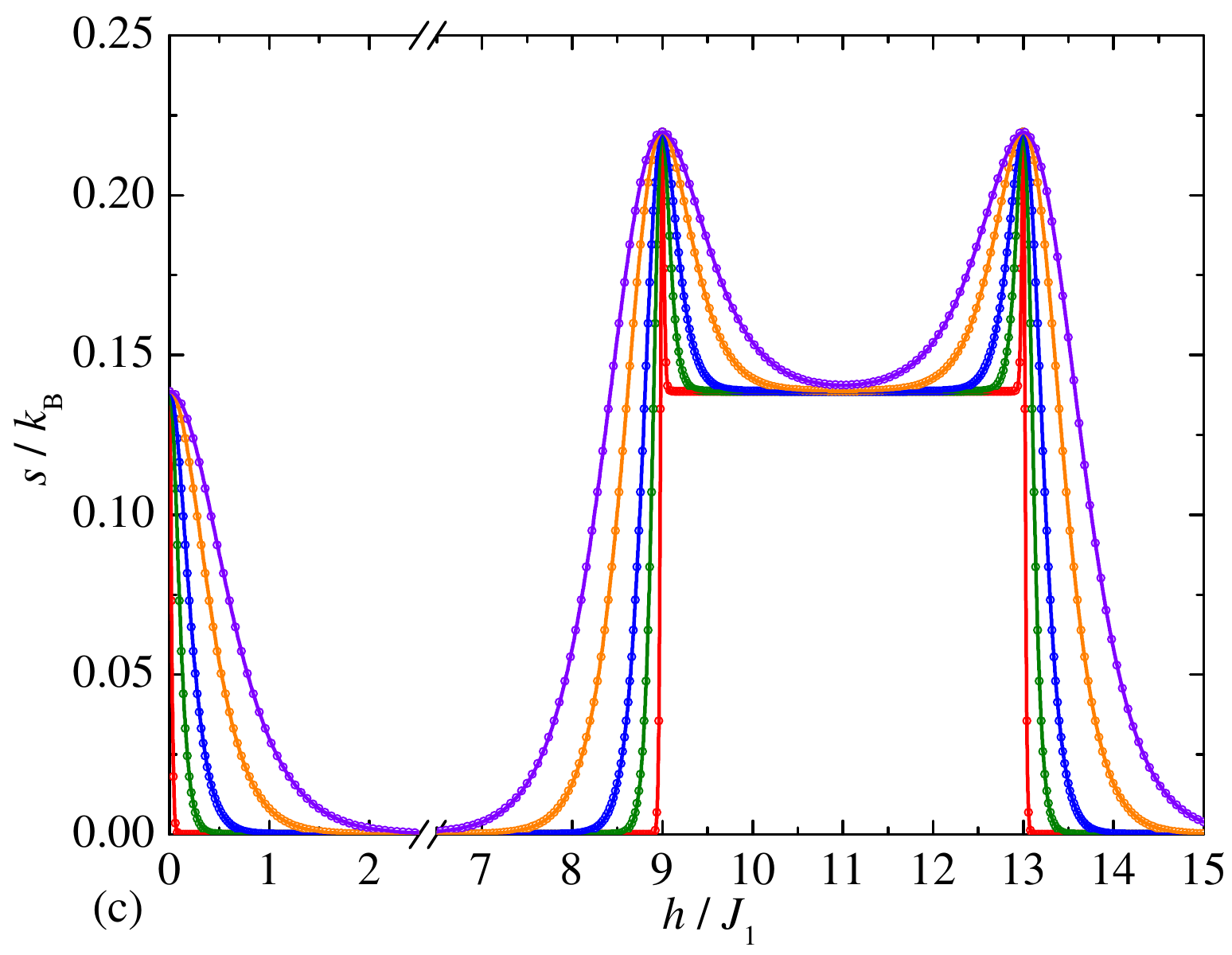}
\hspace{-0.2cm}
\includegraphics[width=0.5\textwidth]{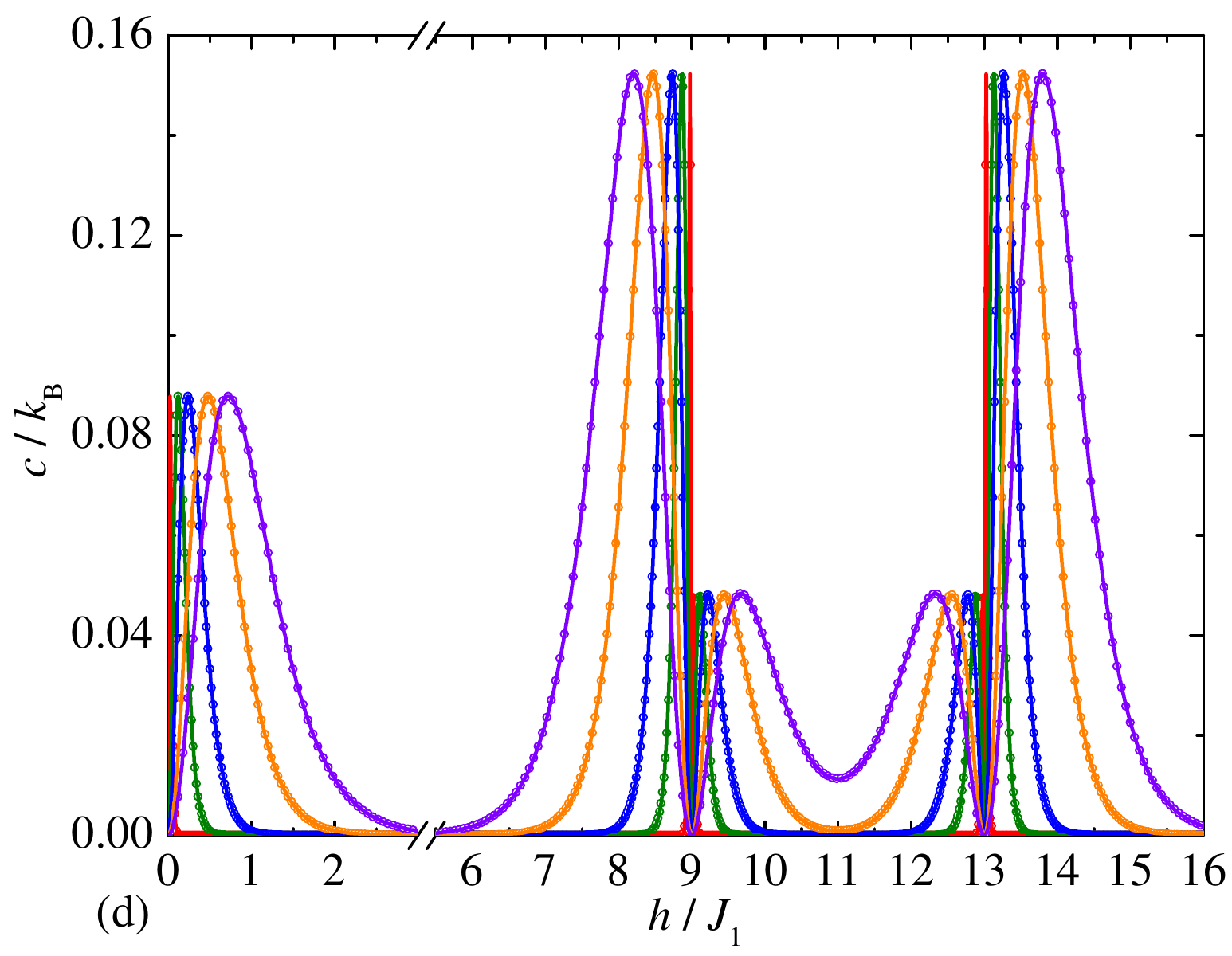}
\end{center}
\vspace{-0.8cm}
\caption{A few isothermal field variations of the magnetization (a), susceptibility (b), entropy (c), and specific heat (d) of the spin-$\frac{1}{2}$ Heisenberg octahedral chain with the interaction ratio $J_2/J_1 = 4$ and $J_3/J_1 = 8$. Symbols show ED data for the finite chain of 20 spins, while solid lines were derived from the analytical expression (\ref{lmgfe}) for the free energy of the effective lattice-gas model in the thermodynamic limit. The legend in the panel (a) applies also to panels (b)-(d).}
\label{fig5}
\end{figure*}

\subsection{Coexistence between bound-magnon plaquette and dimer states}

Next, we turn our attention to the spin-$\frac{1}{2}$ Heisenberg octahedral chain with fixed values of the interaction ratio $J_2/J_1 = 4$ and $J_3/J_1 = 4$ supporting a coexistence of bound plaquette and dimer magnon-crystal phases. In this regime, the bound two- and one-magnon plaquette phases TMP and OMP given by Eqs. (\ref{PS}) and (\ref{OMP}) coexist with the bound two- and one-magnon dimer phases TMD and OMD described by Eqs. (\ref{TMD}) and (\ref{OMD}). Although the magnetization exhibits field dependencies similar to the previous case featuring identical heights of the intermediate plateaus and  positions of the magnetization jumps, there appear several subtle differences reflecting a phase coexistence of the TMP/TMD and OMP/OMD phases as illustrated in Fig. \ref{fig4}(a). A first distinctive feature concerns the temperature-independent crossing points of the low-temperature magnetization curves, which do not occur exactly at the midpoints of the magnetization jumps but instead shift toward the magnetization value of the OMP/OMD phase with the highest degeneracy: $m = 11/50 = 0.22$ at $h/J_1=5$ and $m = 7/20 = 0.35$ at $h/J_1=9$, respectively. Differences in the macroscopic degeneracy of the coexisting bound-magnon crystal ground states are also the primary cause for a temperature-dependent shift of the susceptibility peaks, which move toward the field region associated with the less degenerate magnon-crystal phase  as the temperature increases instead of remaining symmetrically centered around the transition field [see Fig. \ref{fig4}(b)]. These signatures are fully corroborated by the marked field dependence of the entropy shown in Fig. \ref{fig4}(c), which exhibits two identical local maxima $s/k_{\rm B} = \frac{1}{5} \ln 4 \approx 0.2773$ at zero field and at the saturation field $h/J_1 = 9$ in addition to a global maximum  $s/k_{\rm B} = \frac{1}{5} \ln 5 \approx 0.3219$ emerging at the moderate transition field $h/J_1 = 5$. Remarkably, the entropy does not drop to zero even away from the transition fields when it remains at the residual value $s/k_{\rm B} = \frac{1}{5} \ln 2 \approx 0.1386$ in the lower-field region associated with the phase coexistence of TMP/TMD phases and $s/k_{\rm B} = \frac{1}{5} \ln 3 \approx 0.2197$ in the upper-field region corresponding to the phase coexistence of OMP/OMD phases. The significant differences in degeneracy among the competing magnon-crystal phases are also reflected in field variations of the specific heat illustrated in Fig. \ref{fig4}(d), which show a highly asymmetric double-peak profile with sizable difference in peak heights just below and above each transition field. The peak heights $c/k_{\rm B} \approx 0.122$ and $0.0623$ around the first transition field $h/J_1=5$ match the Schottky predictions for two-level systems with relative degeneracies $3:2$ and $2:3$ between the excited and ground state. An even stronger asymmetry in peak heights $c/k_{\rm B} \approx 0.0333$ and $0.2046$ observed around the saturation field $h/J_1=9$ is consistently explained within the Schottky theory for two-level systems with degeneracy ratios $1:3$ and $3:1$.    

\begin{figure}
\begin{center}
\includegraphics[width=0.49\textwidth]{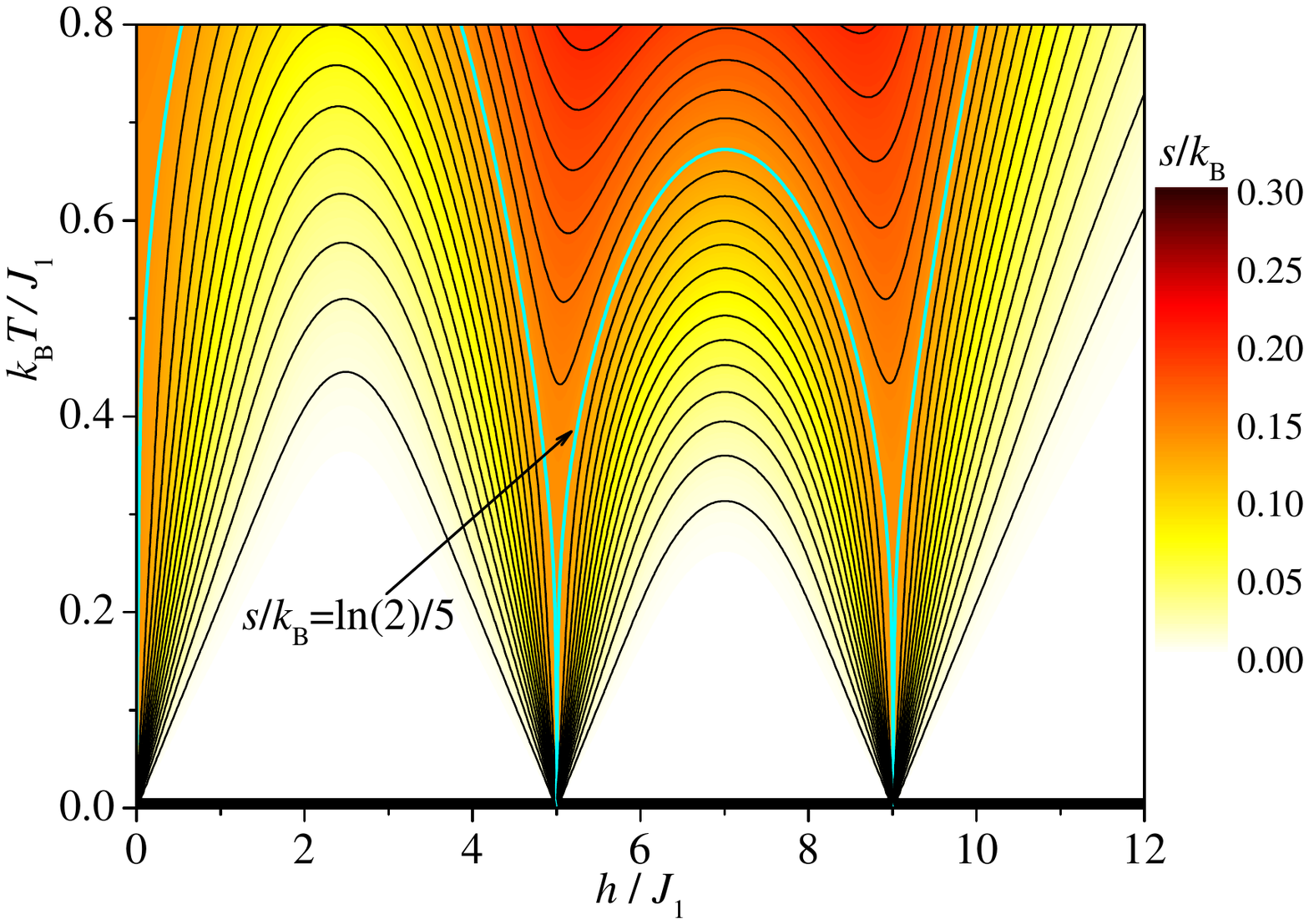}
\includegraphics[width=0.49\textwidth]{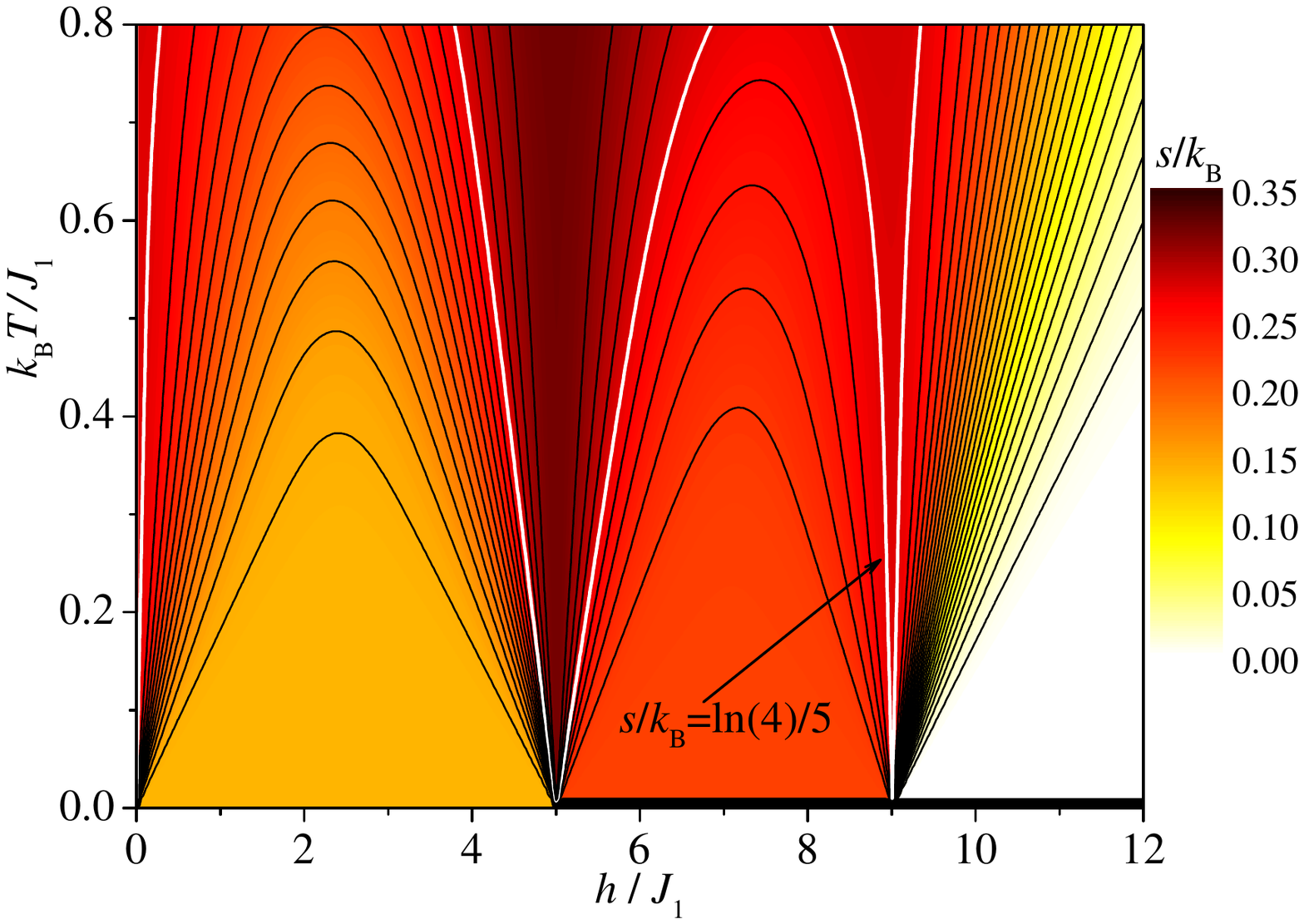}
\includegraphics[width=0.49\textwidth]{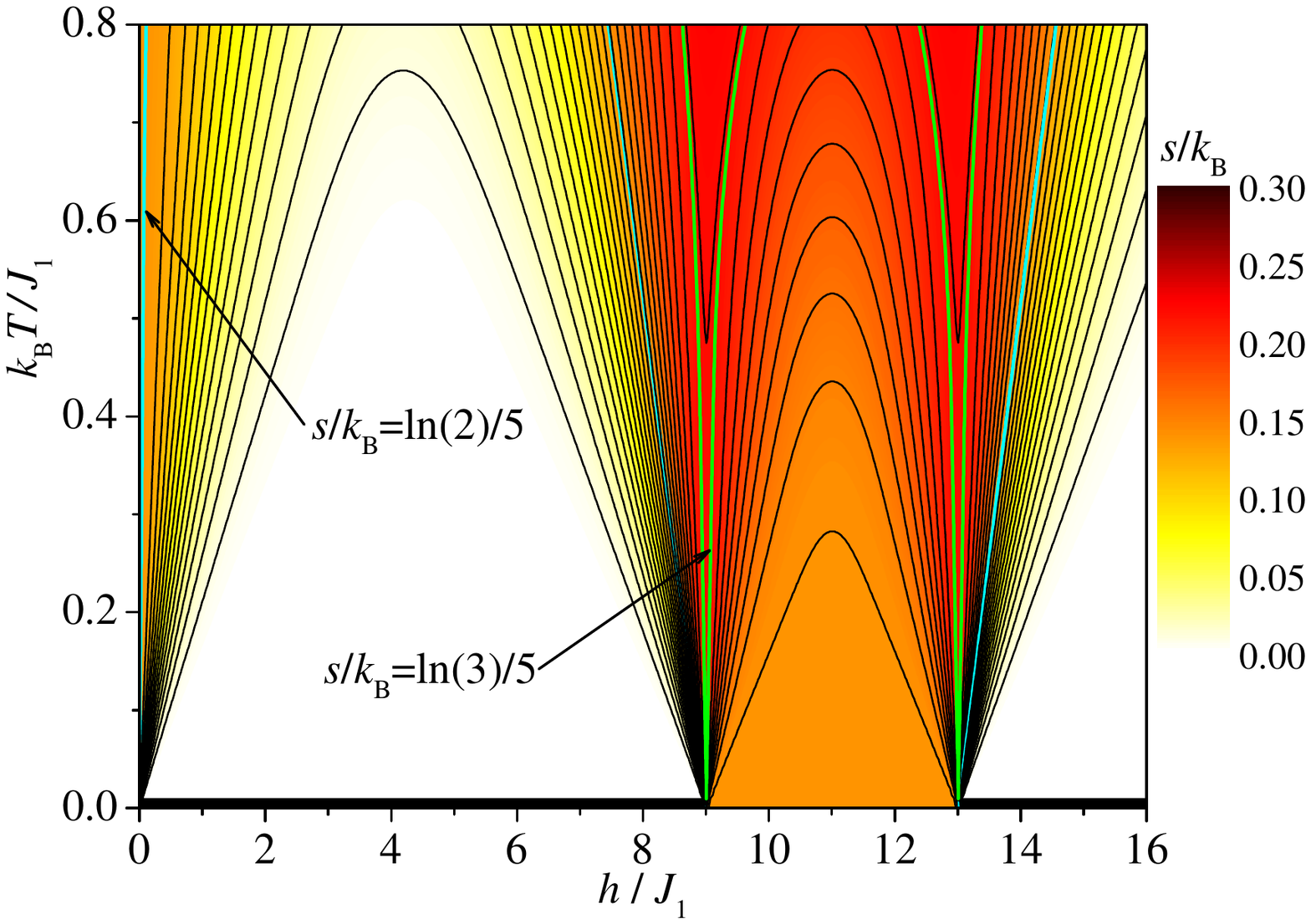}
\end{center}
\vspace{-0.8cm}
\caption{A density plot of the magnetic entropy (per spin) of the spin-$\frac{1}{2}$ Heisenberg octahedral chain in the field-temperature plane as obtained for the effective lattice-gas model in the thermodynamic limit for three representative cases: (a) $J_2/J_1 = 4$, $J_3/J_1 = 2$; (b) $J_2/J_1 = 4$, $J_3/J_1 = 4$; (c) $J_2/J_1 = 4$, $J_3/J_1 = 8$. Black lines are isoentropy contour lines that allow one to trace temperature changes during the adiabatic demagnetization at a fixed value of the magnetic entropy calculated for $s/k_{\rm B} = 0.01, 0.02, \ldots, 0.35$. Light blue, green, and white isoentropy contour lines correspond to the fixed value of the magnetic entropy $s/k_{\rm B} = \frac{1}{5} \ln 2 \approx 0.1386$, $\frac{1}{5} \ln 3 \approx 0.2197$, and $\frac{1}{5} \ln 4 \approx 0.2773$, respectively.}
\label{fig6}
\end{figure}

\subsection{Bound-magnon crystals involving dimer singlets} 

Finally, we examine in detail the spin-$\frac{1}{2}$ Heisenberg octahedral chain at fixed values of the interaction ratio $J_2/J_1 = 4$ and $J_3/J_1 = 8$, which favors the bound two- and one-magnon dimer-singlet phases TMD and OMD as given by Eqs. (\ref{TMD}) and (\ref{OMD}), respectively. It follows from Fig. \ref{fig5}(a) that the dimer-based magnon-crystal phases TMD and OMD give rise to characteristic stepwise magnetization curves with intermediate plateaus at one-fifth and three-fifths of the saturation magnetization. Although the sharp magnetization jumps at field-driven transitions are progressively smoothed with increasing temperature, all low-temperature magnetization curves intersect each other at temperature-independent crossing points with the magnetization values $m = 7/30 \approx 0.2333$ and $11/30 \approx 0.3667$ located at the transition fields $h/J_1 = 9$ and $13$, respectively. The related peaks in the magnetic susceptibility are gradually suppressed as the temperature increases and their positions shift away from the transition fields toward the field region associated with the less degenerate ground state in order to be located further away from the macroscopically degenerate OMD phase [see Fig. \ref{fig5}(b)]. The field variations of the entropy depicted in Fig. \ref{fig5}(c) confirm its nonzero residual value $s/k_{\rm B} = \frac{1}{5} \ln 2 \approx 0.1386$ in the field range delimited by two transition fields $h/J_1 = 9$ and $13$, which is consistent with the macroscopic degeneracy $2^N$ of the OMD phase originating from the two-fold degeneracy of diagonal dimer-singlet bound states on each square plaquette. Outside of this field range, the entropy rapidly drops to zero due to the nondegenerate character of the relevant ground state, while it displays two marked peaks $s/k_{\rm B} = \frac{1}{5} \ln 3 \approx 0.2197$ at both transition fields $h/J_1 = 9$ and $13$ due to a phase coexistence of the respective nondegenerate phase and macroscopically degenerate OMD phase. The peculiar double-peak profile in the specific heat with highly asymmetric peak heights $c/k_{\rm B} \approx 0.0482$ and $0.1524$ shown in Fig. \ref{fig5}(d) can be consistently interpreted within the Schottky theory of a two-level system with relative degeneracies 1:2 and 2:1 between the excited and ground state, respectively.   

\subsection{Enhanced magnetocaloric effect}

Last but not least, we examine the potential of the spin-$\frac{1}{2}$ Heisenberg octahedral chain as a promising working medium for adiabatic refrigeration. Fig. \ref{fig6} presents the density plots of the magnetic entropy of the spin-$\frac{1}{2}$ Heisenberg octahedral chain in the field–temperature plane as obtained from the effective lattice-gas model in the thermodynamic limit for three representative parameter regimes discussed above. Superimposed curves display isoentropy contour lines that trace the temperature evolution during adiabatic demagnetization. 

Fig. \ref{fig6}(a) reveals that the most significant magnetocaloric response occurs for the parameter set supporting the magnon-crystal phases TMP and OMP when the entropy is fixed at $s/k_{\mathrm B} = \frac{1}{5}\ln 2$. This value is consistent with the macroscopic degeneracy of the two-magnon plaquette state TMP at zero field originating from paramagnetic character of the monomer spins, while at the two nonzero transition fields it reflects the phase coexistence of two nondegenerate phases TMP/OMP and OMP/FM, respectively. Under the adiabatic demagnetization along the isoentropy contour line $s/k_{\mathrm B} = \frac{1}{5}\ln 2$, the temperature decreases extremely efficiently and approaches absolute zero with an infinite cooling rate as indicated by the vertical slopes of this contour line both at zero field and at each field-induced transition. For lower entropies $s/k_{\mathrm B} < \frac{1}{5}\ln 2$, adiabatic demagnetization still leads to cooling toward absolute zero temperature, but the slopes of the isoentropy contour lines become finite indicating a reduced yet still substantial magnetocaloric efficiency. For higher entropies $s/k_{\mathrm B} > \frac{1}{5}\ln 2$, the temperature contrarily tends toward a finite value at zero field as well as at both transition fields making the magnetic refrigeration through the adiabatic demagnetization less effective.

Fig.~\ref{fig6}(b) corresponds to the parameter regime in which the bound-magnon plaquette and dimer phases coexist: TMP/TMD coexist at low fields and OMP/OMD at higher fields. Owing to this phase coexistence, the magnetic entropy remains finite throughout the entire field range below the saturation field reflecting the macroscopic degeneracy associated with the competing magnon-crystal states. This persistent nonzero entropy away from the transition fields directly reflects the extensive degeneracy generated by the phase coexistence of TMP/TMD and OMP/OMD magnon crystals and indicates that efficient magnetocaloric cooling can be maintained over a broad range of magnetic fields. The most efficient adiabatic cooling at both zero field and the saturation field is achieved along the contour line with higher entropy $s/k_{\mathrm B} = \frac{1}{5}\ln 4$. At zero field, this entropy value originates from the paramagnetic character of the monomer spins combined with the twofold degeneracy of the singlet states on each square plaquette, whereas at the saturation field it reflects the phase coexistence of the two-fold degenerate OMD state with two nondegenerate OMP and FM states. At intermediate fields, an even larger entropy $s/k_{\mathrm B} = \frac{1}{5}\ln 5$ would be required to reach a similarly divergent cooling rate although this particular contour line is not shown in the figure. 

In the last parameter regime illustrated in Fig.~\ref{fig6}(c), the intermediate-field region is governed by the macroscopically degenerate OMD phase, whereas the low-field region hosts the nondegenerate TMD phase, becoming the macroscopically degenerate exactly at zero field due to the paramagnetic character of the monomer spins. In this case, the most efficient adiabatic cooling with a divergent cooling rate is achieved at both finite transition fields along the isoentropy contour line $s/k_{\mathrm B} = \frac{1}{5}\ln 3$, while at zero field the optimal cooling occurs at the lower entropy $s/k_{\mathrm B} = \frac{1}{5}\ln 2$ consistent with the $2^N$-fold degeneracy of the TMD phase. A particularly remarkable feature emerges for entropies below $s/k_{\mathrm B} < \frac{1}{5}\ln 2$: adiabatic demagnetization first cools the system to absolute zero already at the saturation field, after which the temperature remains pinned at absolute zero temperature throughout the entire field interval $9 < h/J_1 < 13$ associated with the massively degenerate OMD phase. Only upon entering the lower-field TMD region $h/J_1 < 9$ does the temperature begin to rise before it ultimately decreases again toward zero as the magnetic field is completely removed adiabatically. However, the isoentropy contour lines  approach absolute zero temperature at zero field always with a finite cooling rate (slope) for any lower entropy $s/k_{\mathrm B} < \frac{1}{5}\ln 2$ in contrast to the divergent cooling rate obtained for $s/k_{\mathrm B} = \frac{1}{5}\ln 2$. For higher entropies $s/k_{\mathrm B} > \frac{1}{5}\ln 2$, the temperature tends toward a finite value at zero field making adiabatic demagnetization significantly less effective in this regime.

\section{Conclusion}
\label{sec:conc}

In this work, we have carried out a comprehensive analytical and numerical investigation of the spin-$\frac{1}{2}$ Heisenberg octahedral chain in the highly frustrated parameter regime. By combining the variational principle and the concept of independent bound magnons we have developed within an extended localized-magnon theory a five-component lattice-gas model, which enabled us to identify several bound-magnon crystal ground states and to elucidate the interplay between frustration, magnetic field, thermal and quantum fluctuations.

In a low-field region, we rigorously established two fragmented bound-magnon crystal phases involving either the plaquette-singlet state or the product state of two dimer singlets on each square plaquette. In a high-field region, we rigorously proved the existence of two additional bound-magnon crystal phases involving either a collective one-magnon plaquette state or a single dimer-singlet state on each square plaquette. Full ED of finite chains provided excellent quantitative agreement with the predictions of the extended localized-magnon theory and confirmed the microscopic nature of all four bound-magnon crystal phases.

The low-temperature features were examined in detail for three representative parameter regimes revealing striking magnetization plateaus, characteristic anomalies in the susceptibility and specific heat, and field-induced entropy peaks directly related to macroscopic degeneracies of the underlying bound magnon-crystal phases. The extended lattice-gas model accurately reproduced all finite-temperature features and allowed us to interpret each anomaly in terms of the competition between distinct bound-magnon crystal states. In particular, we demonstrated that the location of temperature-independent crossing points in the low-temperature magnetization curves and the relative peak heights in field dependencies of the specific heat may unveil the degeneracy of these bound-magnon crystal phases.

Finally, we furnished evidence that the spin-$\frac{1}{2}$ Heisenberg octahedral chain exhibits an exceptionally rich magnetocaloric response. In the highly frustrated regime, the system supports efficient adiabatic cooling not only near zero field and at field-driven transitions but also across extended field intervals associated with a macroscopically degenerate magnon crystal. In particular, we identified optimal entropy values at which isoentropy lines develop vertical tangents signaling an infinite cooling rate and uncovered unusual scenarios in which the temperature becomes pinned at absolute zero over a wide field range. These findings may stimulate further exploration of frustrated quantum magnets as promising candidates for efficient magnetic refrigeration technologies and motivate the search for experimental realizations of the octahedral chain geometry.

\begin{acknowledgments}
This work was funded by the grant of The Ministry of Education, Research, Development and Youth of the Slovak Republic under the contract No. VEGA 1/0298/25 and by the grant of the Slovak Research and Development Agency under the contract No. APVV-24-0091. J.S. would like to dedicate the present paper to the memory of Johannes Richter, who touched his life through his kindness and continued support. He gratefully acknowledges the many inspiring discussions they shared, Johannes’s insightful ideas, and his readiness to help whenever needed. J.S. also wishes to recognize his kind hospitality in Magdeburg and Dresden, as well as, his friendship, which will be remembered with sincere appreciation.

\end{acknowledgments}

\end{document}